\documentclass[usenatbib,twocolumn]{mn2e}
\usepackage{graphicx}
\usepackage{amssymb}
\usepackage{subfigure}
\newcommand{\beq}{\begin{equation}}
\newcommand{\eeq}{\end{equation}}

\def\mA{\mathcal A}

\def\beq{\begin{equation}}
\def\ee{\end{equation}}
\def\lsim{\mathrel{\rlap{\lower4pt\hbox{\hskip1pt$\sim$}}
    \raise1pt\hbox{$<$}}}
\def\gsim{\mathrel{\rlap{\lower4pt\hbox{\hskip1pt$\sim$}}
    \raise1pt\hbox{$>$}}}
\def\bfE{{\bf E}}

\def\bfB{{\bf B}}

\def\ts{\times}
\def\lb{\langle}
\def\rb{\rangle}
\def\curl{\nabla {\ts}}

\def\bfv{{\bf v}}

\def\bfb{{\bf b}}

\def\bfB{{\bf B}}

\title[Activity, rotation, mass-loss, and B-field]
 {Minimalist coupled  evolution  model for  
 stellar  x-ray activity, rotation, mass loss, and magnetic field}
 
\author [Blackman \& Owen]
{Eric G. Blackman$^{1,2,3}$\thanks{E-mail: blackman@pas.rochester.edu}\thanks{IBM-Einstein Fellow/Simons Fellow}, James E. Owen,
$^{3}$\thanks{E-mail: jowen@ias.edu}\thanks{Hubble Fellow}\\ 
 $^{1}$Department of Physics and Astronomy, University of Rochester, Rochester NY, 14627, USA\\
 $^{2}$Laboratory for Laser Energetics,  University of Rochester, Rochester NY, 14623, USA\\
 $^{3}$School of Natural Sciences, Institute for Advanced Study, Princeton NJ, 08540 USA\\ }

\begin{document}

\date{}
\pagerange{\pageref{firstpage}--\pageref{lastpage}} \pubyear{}
\maketitle
\label{firstpage}
\begin{abstract}
Late-type  main sequence stars exhibit an x-ray to bolometric flux ratio that depends on ${\tilde Ro}$, the ratio of rotation period to  convective turnover time,  as  ${\tilde Ro}^{-\zeta}$ with $2\le \zeta \le 3$ for ${\tilde Ro} >0.13$,  but saturates with $|\zeta| <0.2$  for ${\tilde Ro} < 0.13$.  Saturated stars are younger than unsaturated stars and show a broader spread of rotation rates and x-ray activity. The unsaturated stars  have magnetic fields and rotation speeds that scale roughly with the square root of their age, though possibly flattening for stars older than the sun.  The  connection between faster rotators, stronger fields, and higher activity has been  established observationally, but a  theory for the unified time-evolution of x-ray luminosity, rotation, magnetic field and mass loss that captures the above trends has been  lacking.  Here we derive a minimalist holistic framework for the time evolution of these quantities built from  combining a Parker wind with new ingredients: (1) explicit sourcing of both the thermal energy launching the wind and  the x-ray luminosity via dynamo produced magnetic fields; (2)  explicit coupling of x-ray activity and mass loss saturation to  dynamo saturation (via magnetic helicity build-up and convection eddy shredding); (3) use  of  coronal equilibrium to determine how magnetic energy is divided into wind and x-ray contributions.   For solar-type stars younger than the sun, we infer conduction to be a subdominant power loss compared to x-rays and wind.  For older stars,  conduction is more important, possibly quenching the wind and reducing angular momentum loss.  We focus on the time evolution for stars  younger than the sun, highlighting what is possible for further generalizations. Overall,   the approach shows promise toward a unified  explanation of all of the aforementioned observational trends.
\end{abstract}


\section{Introduction}

Understanding  the mutual evolution of x-ray activity, magnetic fields, rotation, and spots in stars
 comprises a rich enterprise of research not only for the basic astrophysics 
 but  for bolstering  gyro-chronological methods  of stellar ageing \citep{Skumanich1972,Mamajek2008,Epstein2014} and  gauging the influence of such activity  on  the atmospheres and/or habitability zones of companion planets \citep{Owen2013,Owen2014,Tarduno2014}. 
Given that the number of stars with rotation and variability measurements has now increased by more than  an order of magnitude with COROT and Kepler data \citep{Gilliland2010,McQuillan2014}
 further developments in theoretical work are  timely. 

%
 
Observed relations between  coronal activity (measured as the ratio of x-ray to bolometric luminosity  $
R_x\equiv
 {\mathcal{L}_x/\mathcal{L}_{bol}}$) and rotation period of main sequence F,G,K and M (earlier than M3.5) stars show that \citep{Pallavicini1981,Noyes1984,Vilhu1984,Micela1985,Randich2000,Montesinos2001,Pizzolato2003,Wright2011,Vidotto2014,Reiners2014},
\beq
{\mathcal{L}_x\over \mathcal{L}_{bol}} \propto {\tilde Ro}^{-\zeta},
\label{rossby1}
\eeq
where the Rossby number ${\tilde Ro}$ is usually defined such that
${\tilde Ro} \equiv 2\pi Co^{-1}\equiv {2\pi / \Omega \tau_c} ={P/ \tau_c}$,  where $Co$ is the Coriolis number,
 $\Omega$ is the surface angular velocity, $P$  is the associated rotation period, and $\tau_c$ is the convective turnover time.
For  the ``unsaturated'' regime of ${\tilde Ro}>0.13$, 
  the data show that $2\le \zeta \le 3$ , whilst for the ``saturated'' regime  ${\tilde Ro} <0.13$, the data show $|\zeta|<0.2$
   \citep{Wright2011,Reiners2014}.
The rotation period is measured  directly from time-series photometry of variability associated  with star spots,  but the value of 
$\tau_c$  is  inferred from stellar models, by matching to
a given colour index \citep{Noyes1984}.

Direct spectropolarimetric stellar observations  along with  solar observations and theory linking particle energization  to sites of magnetic energy dissipation \citep{Schrijver2000,Vidotto2014}  have long indicated that stellar magnetic field strength correlates with  x-ray activity.
    For the sun,  the solar cycle of  activity  is  correlated with flux sign reversals of the large scale field.  This highlights that the field must be amplified  by  internal dynamo action which is
     buoyantly sourcing the corona, not  merely a vestige of flux freezing from pre-main sequence evolution. 
  Activity-amplitude cycles  have long been  observed in many  stars \citep{Baliunas1998}, while 
  recent observations are  starting to  reveal  large scale stellar field reversals as well
   \citep{Morgenthaler2011}.
   
The link between  activity and dynamos  implies that  some combination of thermal, rotational, and differential rotational  energy sources the magnetic fields
 \citep{Moffatt1978,Parker1979,Krause1980,Christensen2009,Blackman2015}. 
 The  qualitative trend for  faster rotators to have  larger $R_x$ in the unsaturated regime 
 suggests that  dynamos produce stronger fields for faster rotators 
  \citep{Noyes1984,Montesinos2001,Wright2011}. However, theoretical
  scenarios quantitatively connecting the saturated regime of $R_x$  to the physics of  dynamo saturation have only begun to  emerge recently \cite{Blackman2015,Pipin2015}.

Since  large scale fields produced by dynamos can transport angular momentum
in stellar winds, older stars are expected to be  slower rotators. As mentioned, this  trend is evident for low mass main sequence stars via  the  Skumanich relation $\Omega_* \propto t^{0.55}$  
 \citep{Skumanich1972,Mamajek2014} for stars in the saturated regime younger than the sun,
 though the scaling seems to flatten for older stars (\cite{vanSaders2016}).
 In addition, 
 the wind mass loss rate $\dot M$ measured from line emission associated with wind-ambient media interactions   \citep{Wood2005,Wood2014}, is also correlated with x-ray activity  in the form
${\dot M}\propto F_x^{1.34 \pm 0.18}$ for stars with x-ray fluxes $F_X< 8\times 10^5\ {\rm erg \ cm^{-2}s^{-1}}$, highlighting that mass loss  also evolves with time.
The large scale magnetic field also follows a similar trend with the surface averaged radial field declining as $|B_r|\propto t^{-0.655\pm 0.045}$ and  scales similarly to the  magnetic  flux,  $|\Phi_r|\propto t^{-0.622\pm 0.042}$ \citep{Vidotto2014}.
The unsaturated rotators for which these relations apply, likely evolved from
rotators originally in the saturated regime for which the observed spread in rotation
rates over stellar populations are  much larger than in the saturated regime \citep{Gallet2013}.


While the observations  linking time evolution of x-ray activity, rotation, and magnetic fields with age are improving, there has yet to be a time-dependent theory that captures all of the observed scalings from basic  principles. Assuming that the stellar wind is {\it initially} energised\footnote{We are explicitly ignoring heating that may occur at higher heights on the way to the sonic point as we shall discuss later.} in the corona,  the mechanical luminosity associated with wind launching is then $L_{\rm mech}=\dot{M}_\odot c_VT_{\rm x\odot}\approx1\times10^{27}$~erg s$^{-1}$, 
where $c_V=3k_B/m$ is the heat capacity at constant volume for a plasma with 
mean molecular weight $1/2$ and proton mass $m_p$, and Boltzmann constant $k_B$.
Comparing this to the Sun's x-ray luminosity ($L_{x\odot}\approx2\times10^{27}$~erg s$^{-1}$) we note  that they are of similar order. Since the magnetic field is the energy source for both the x-ray emission and the wind, treating the evolution of the x-ray 
luminosity and the rotational evolution independently would be inconsistent. 
There is no {\it a priori} reason why the x-ray luminosity and mechanical luminosity of the wind should scale similarly with time, so only a coupled solution to the x-ray activity, rotation and magnetic field evolution can provide a correct description of  x-ray activity and rotational evolution of a star.    

Here we pursue a minimalist holistic model  for the coupled time evolution of the x-ray luminosity, rotation, mass loss, and magnetic field strength as a basis for further theoretical exploration.   We consider that a dynamo supplies  magnetic field to the corona, some fraction of this field dissipates in  closed field lines and  is a source of the thermal x-ray emitting gas.
Some of the field lines open up, allowing the hot gas to propagate along them  in a  Parker-like stellar wind.
  The magnetic field in this wind extracts angular momentum from the star, which in turn reduces the dynamo-produced field strength, along with the x-ray luminosity, and wind mass supply.  The key  new ingredients in our approach  are:  
  1) making use of the assumption that (closed)  dissipation of field produced by the dynamo 
  sources  both the thermal plasma, x-ray luminosity, and mass outflow rate;  2) explicit  coupling of  the magnetic field evolution to the other dynamical variables using a physically motivated dynamo saturation model which  captures the $\tilde Ro$  dependence of stellar x-ray luminosities based on \citep{Blackman2015}; 
  (3) use of a coronal equilibrium model  to determine how the magnetic energy sourcing divides into wind and x-rays.
  
Our  to the approach to the wind momentum evolution differs from   complementary semi-empirical approaches  \citep{Matt2012,Reiners2012,vanSaders2013,Gallet2013,Matt2015} for which 
MHD simulations are taken to empirically inform an exponent  parameter  that 
 determines how the spin-down torque depends on magnetic field strength, mass loss rate, 
 mass, and
 radius.  These approaches  leave a degeneracy between wind mass loss rate and magnetic field and make   an empirical choice to close this relation.  
 Although we use a more idealized standard  torque expression from  an equatorial wind,  our expression for  magnetic field comes from   dynamo saturation arguments  and we eliminate the degeneracy between magnetic energy source origin,  mass outflow rate, and x-ray luminosity  via  consideration of the additional physics of coronal equilibria supplied by magnetic energy.
 
 Our approach also differs from    \cite{Cranmer2011} and  \cite{Suzuki2013} who 
 focus respectively   on thoughtful   semi-analytic and numerical magnetohydrodynamic models of 
  Alfv\'en wave driving of stellar wind mass loss.    These papers  focus on mass loss as function of imposed field strength and filling fraction
(using plausible choices and considerations) but 
 do not solve for  the coupled  time evolution  of spin, magnetic field, mass loss and luminosity.

In Sec. 2 we summarize the  basic Parker wind solution. 
In Sec. 3 we provide the relation between magnetic field, Rossby number and dynamo theory.  In Sec. 4 we use the  coronal flux of  dynamo-produced magnetic field as an energy source for the  x-ray emission and  mass loss. We  obtain an expression for x-ray luminosity and
mass loss rate as a function of the equilibrium coronal temperature  which is consistent with observations, and is later needed in Sec. 6.
In Sec. 5 we derive the needed equations for time evolution of angular momentum and
toroidal coronal magnetic field. In Sec. 6, we combine the  results of the previous sections to
solve for the time-evolution of  x-ray luminosity, mass loss, magnetic field, and rotation rate from a magnetically driven Parker wind, discussing the results in the context of observations. 
We conclude in Sec. 7. 

\section{Radial  Wind and Solar Properties}

\subsection{Parker Wind solution}
For simplicity, we employ an isothermal Parker wind solution for the radial velocity \citep{Parker1958} and in the next sections
 connect the magnetic heating source to the wind temperature,  x-ray luminosity, and stellar spin-down.

We assume that the time scale for the wind solution to evolve at fixed radius $r$ is long compared to the wind propagation time  from the star to  $r$, so we  use the steady-state wind solutions
and consider their secular evolution later.
The continuity equation for a  spherical wind then gives 
\beq
{\dot M}=4\pi r^2 \rho U_r,
\label{mdotdef}
\eeq
where $\rho$ and $U_r$ are the radial dependent density and radial wind speed respectively.
The radial momentum  equation for a spherical wind, assuming that the toroidal magnetic pressure along field lines  is small, is
\beq
{1\over 2}\left(1-{c_s^2\over U_r^2}\right){d U_r^2\over dr}=-{GM\over r^2 }\left(1-{2c_s^2r \over  GM}\right),
\label{2.2}
\eeq
where the isothermal sound speed, $c_s \propto T^{1/2}$ and $T$ is the coronal x-ray temperature.
At  $U_r=c_s$, $r=r_s$ where the sonic radius is given by 
\beq
{r_s\over r_*}={GM\over 2c_s^2r_*}.
\label{2.4}
\eeq
The general solution of Eq. (\ref{2.2}) is
\beq
{U_r^2\over c_s^2}-ln\left({U_r^2\over c_s^2}\right)=4 ln {r\over r_s} +4 {r\over r_s} -3.
\eeq
For $r<<r_s$, this reduces to 
\beq
U_r(r)=c_s e^{3/2}e^{-2(r_s/r)}
\label{2.6}
\eeq
and  for
 $r>>r_s$
the solution is
\beq
U_r(r)=2 c_s [ln (r/r_s)]^{1/2},
\label{2.7}
\eeq
which are  standard results \citep{Parker1958}.

\subsection{Solar values as   scaling parameters}

We compile some solar  quantitites here for later use.

For the solar radius, mass,  moment of inertia, and age  we take   \citep{Cox2000} 
$r_*=R_\odot=6.955\times 10^{10}$; $M_\odot=2\times 10^{33}$g; 
 ${\mathcal I}_{\odot}=0.059 M_\odot R_\odot^2\ {\rm g \cdot cm^2}$; and 
 $t_{\odot}=4.6\times 10^9$yr  respectively.
 
 For the average solar x-ray luminosity  and magnetic properties we take \citep{Aschwanden2004} 
 ${\mathcal L}_{x\odot}=6\times 10^{-7}{\mathcal L}_\odot$ erg/s, where ${\mathcal L}_\odot=4\times 10^{33}$erg/s.  
 The average solar coronal x-ray temperature is   $T_{x\odot}\simeq 1.5\times 10^6$K although we will explicitly explore  consequences of other choices in sections
4.1 and 6.3.
 We assume a surface radial field of   $B_{r\odot}=2$G,  surface toroidal field $B_{\phi\odot}=1.56\times 10^{-2}$G,     rotation speed
   $\Omega_\odot=2.97 \times 10^{-6}/{\rm s}^2$
  and ${\tilde Ro}_\odot=2$.

 For the solar wind  \citep{Cranmer2012} we use a mass loss rate
 ${\dot M}_\odot=1.3\times 10^{12}$g/s; 
sonic radius of
   $r_{s\odot}=2.58\times 10^{11}$cm; and
  Alfv\'en radius of  $r_{A\odot}=1.63\times 10^{12}$cm.  The  associated  density and  outflow speed for the latter   are
 $\rho_{A\odot}\equiv \rho(r_A)=1.51\times 10^{-21}\ {\rm g/cm^3}$
and
  $u_{A\odot}=2.64\times 10^7$cm/s.

\section{Magnetic Field and Rossby Number}

Dynamo theorists
have  augmented  20th-century textbook mean field  theory to include  a tracking of the evolution of magnetic helicity  (for reviews see \citet{Brandenburg2005,Blackman2015H}).  
Though  still a very active area of research, a takeaway improvement to textbook stellar dynamos is that  the ``$\alpha$'' effect, which represents the pseudoscalar coefficient of the turbulent electromotive force along the mean magnetic field, is best represented  as the difference  $\alpha_0-\alpha_M$, where 
 $\alpha_M=\lb\bfb\cdot \curl \bfb\rb\tau_{ed}$,  is
proportional to the  current helicity density of magnetic fluctuations and \citep{Durney1982}
\beq
\alpha_0\simeq {\tau_{ed}\over 3} \lb\bfv\cdot\curl\bfv \rb \sim {q_\alpha\over 6}\tau_{ed}^2 {\Omega v^2\over r_c} cos \theta_s
\label{alpha0}
\eeq
is the kinetic helicity (the usual ``Parker $\alpha$-effect''). Here  $\Omega_*$ is 
the surface rotation speed; $\theta_s$ is a fiducial polar angle; $v$ is a typical turbulent convective velocity;  $\tau_{ed}$ is the correlation time of the turbulence for radial
motion;  $0<q_\alpha<1$ is a product of terms accounting for  anisotropy in the convection and the ratio $\Omega_*/\Omega(r_c)$, with  $r_c$  a fiducial radius in the convection
zone (taken as its base for the sun). 

Blackman \& Thomas (2015) argued that  turbulent correlation time entering the dynamo coefficients $\tau_{ed}$ should equal the convection time $\tau_c$ for ${\tilde Ro}>>1$, but equal the shear time scale from internal differential rotation for ${\tilde Ro}<<1$ as rapid shear would  shred eddies rapidly in the latter limit. To capture these regimes \cite{Blackman2015} write:
\beq
\tau_{ed}= {sP\over 1+  s {\tilde Ro}}.
\label{taued}
\eeq
where $s$ is a shear parameter defined such that 
$|\Omega_0-\Omega(r_c,\theta_s)|=\Omega_0/s$, where $\Omega_0=\Omega(r_*,\theta_s)$ is the surface rotation speed. \footnote{Even in the absence of shear,
 strong rotation turns convection cells into cylindrical rolls and the heat transport
 scale may be reduced while the eddy time remains  constant \citep{Barker2014}. Other prescriptions for the influence of rotation on 
 the effective turbulent eddy and time scales  therefore warrant investigation.}
  
  

The  importance of the difference $\alpha_0-\alpha_M$ was evident in the spectral  approach of 
\cite{Pouquet1976} but  is made more conspicuous in a  two-scale mean-field  approach  \cite{Blackman2002}. In the Coulomb gauge, 
or in an arbitrary gauge when the  averaging scale significantly exceeds the fluctuation scale  \citep{Subramanian2006},    
$\alpha_M$ is proportional to the magnetic helicity density of fluctuations.
\cite{Blackman2003b} argue that 
stellar dynamos at large magnetic Reynolds numbers
 might saturate as 
$\alpha_0$ drives  large-scale helical magnetic  field growth, but magnetic helicity conservation
leads to a build up of $\alpha_M$  that nearly offsets $\alpha_0$. 
The field would decay were it not for helicity fluxes 
 of the small scale fluctuations 
 through the star that sustains the mean electromotive force that in turns
 sustains the dynamo (see also  \cite{Shukurov2006}).

 The large scale poloidal field strength in this circumstance is
 approximately equal to that of the helical field whose value
is estimated by setting  $\alpha_0-\alpha_M\simeq 0$ and using magnetic helicity conservation to connect $\alpha_M$ to the large scale helical field strength.  The toroidal field is  amplified non-helically  by differential rotation. Downward 
turbulent pumping \citep{Tobias2001} hampers buoyant loss, but only above a threshold field strength \citep{Weber2013,Mitra2014} such that $\alpha_M$ can still  approach the value  $\alpha_0$ before buoyancy kicks in.
A saturated  dynamo is  then maintained with large-field amplification balanced by   buoyant loss, itself coupled to the beneficial 
loss of small scale helicity.

Following \cite{Blackman2015}, the saturated large-scale poloidal field inside the convection zone 
based on the aforementioned $\alpha_M \sim \alpha_0$ is  
\beq
B_p^2 \sim  8\pi{l_{ed}\over L_\alpha} f_h \rho  v^2, 
\label{bpol}
\eeq 
where $l_{ed}/L_\alpha$ is the ratio of convective eddy scale to the thickness of the zone in which $\alpha$ operates and $f_h$ is  the fractional helicity 
given by:
\begin{equation}
f_h= {l_{ed}|\lb \bfv \cdot \curl \bfv\rb| / v^2} = {q_\alpha cos \theta_s \over 6}{s\over 1+  s {\tilde Ro}}{l_{ed}\over r_c}.
\end{equation}
The  toroidal field is linearly amplified by shear above this value during a buoyant loss time 
$\tau_b\sim L_\alpha/u_b$, where $u_b$ is a typical buoyancy speed for those structures that escape. Thus the total field satisfies
\beq
{B^2\over 8\pi} \simeq 
 {B_p ^2\over 8{\rm \pi}}(1+\Omega \tau_b/s)^2\sim
 {B_p^2\over 8\pi}(\Omega \tau_b/s)^2,
\label{16}
\eeq
where the latter similarity applies for  $B_\phi > B_p$ 
 and is valid for  $\Omega \tau_b /s>1$, which applies even for slow rotators like the sun (Blackman \& Thomas 2015). If we take
$u_b\simeq {B_\phi^2/(12 \pi \rho v)}$ for the rise of buoyant flux 
 tubes (Parker 1979; Moreno-Insertis 1986; Vishniac 1995; Weber et al. 2011), 
  then $\tau_b\simeq{ 12\pi L_\alpha  \rho v/ B_\phi^2}$. 
 Using these in  (\ref{16}) and solving for $B^2\sim B_\phi^2$   gives 
\beq
\begin{array}{r}
{B^2\over 8\pi}\simeq   
  {(3\pi)^{2\over 3}} \left({L_\alpha q_\alpha cos\theta\over 6r_c}\right)^{1\over 3}\rho v^2
 {s^{1/3}\over (1+ s {\tilde R}_o)},
\label{bsat}
\end{array}
\eeq 
using the above expression for $f_h$, Eq. (\ref{bpol}),   $\tau_{ed}\Omega = {2\pi l_{ed}\over v P}$ and
Eq. (\ref{taued}). 
Eq. (\ref{bsat}) also  agrees with the   ${\tilde Ro}<<1$  scaling of Christensen et al. (2009) 
in that $B^2$ becomes independent of $\Omega$, and  $B^2\propto \rho^{1/3} ({\rho v^3})^{2/3}$.

As this  primarily toroidal interior field rises to the surface,
it transforms into  poloidal loops that interact and produce a net poloidal field in each hemisphere.
 We  therefore take Eq. (\ref{bsat}) to scale with the radial  surface mean field.
 We hereafter use subscript $r$ to indicate surface poloidal field magnitude.
 We assume  the dominant variable  on the right hand side of   Eq. (\ref{bsat}) is $\Omega=2\pi/P$ (and thus $\tilde Ro$)  and normalize most all dynamical quantities   to values of the present day sun given in section 2.2.
We  extract from  Eq. (\ref{bsat})  
the normalized surface radial magnetic field magnitude 
\beq
b_r\equiv g_L(t) {B_{r*}(t)\over B_{r\odot}}=g_L(t)\left({s \over s_\odot }\right)^{1/6}
{\sqrt{1+ s_\odot {\tilde Ro}_\odot} \over  \sqrt{1+ s {\tilde Ro}}},
\label{2.42}
\eeq
where the function $g_L(t)$  is dervied later (below Eq. \ref{lx}) and deviates from unity only if  ${\mathcal L}_{bol}$ evolves.



\section{ Magnetic Energy as Source of X-ray luminosity and Outflow}
 
We can make substantial progress toward a global evolution model 
by  positing  that the thermal energy driving the wind from the corona and the coronal x-ray luminosity  both result from  dynamo produced fields. 
We assume that the hot x-ray emitting coronal gas is the same hot gas at the base of the wind.  
We  physically motivate  this simplifying assumption  by noting that   much of the plasma which becomes the solar wind gets injected onto
open field lines in the dynamical opening of  closed field lines during reconnection events. This   blurs the distinction between
the plasma source which supplies the wind and that which accounts for the x-ray temperature, as long as the cooling time is long.

Some previous models \footnote{We thank the referee for references cited in this paragraph.}
 have also assumed the equality of these two temperatures 
\citep{Parker1958,Pneuman1971,Holzwarth2007,Vidotto2012}
Such  models then result in a wind speed that is  dependent on the the temperature at the base of the wind.
Alternatively, by demanding a specific  terminal wind speed, models can also be constructed which  then extract the required wind base  temperature a posteriori.
\citep{Matt2008,Cranmer2011,Johnstone2015}.
In our framework, the simplest way to incorporate a different coronal and wind temperature at the base 
would be to assume a prescribed functional form
that relates the two.  This could be carried through our calculations without practical difficulty,  but 
but would add an additional  function which we choose to avoid.
As there is no clear dynamical prescription to relate the coronal temperature to that at the base of the wind 
we proceed to assume the coronal temperature is the same as that at the wind base,
leaving other options for future work.


 \subsection{Coronal equilibrium determines the relation between ${\mathcal L}_x$, $\dot M$, and $T$}
 
On  time scales short with respect to  stellar spin-down and averaged over
multiple cycle periods, we consider the x-ray emitting corona as the region one density 
scale height above the chromosphere. We assume that it is in equilibrium, balancing
 magnetically driven heating by losses from radiation,  conduction, and outflow.
To determine the equilibrium temperature and thermal pressure, we follow a similar procedure to  \cite{Hearn1975} but  focus on the balance within one scale height.  

For a temperature-independent heating source,  the  temperature derivative of the energy balance
equation at constant pressure is
 \beq
  {\partial \over \partial T} (F_{W1}+ F_{x} + F_{C})=0,
  \label{equ}
 \eeq
 where $F_{W1}$ is the wind flux from the single scale height, and 
 $F_x$ and $F_C$ are the radiative (x-ray) and conductive loss from that  region.
We now give expressions for $F_{W1}$, $F_x$ and $F_C$.

The energy per unit time advected away by  mass loss within a coronal scale height is dominated by 
 thermal energy and is given by 
 \beq
 F_{W1}={{\dot M}\over 4\pi R_0^2}c_V T=1.5{{\dot M}\over 4\pi R_0^2}c_s^2 
=1.5 p_0 c_s e^{3/2}e^{-{3.9\over {\tilde T}}{m_*\over r_*}},
\label{14}
 \eeq
   where $\tilde T= {T\over 3 \times 10^6{\rm K}}$; 
 $c_V=3k_B/m_H$ is the heat capacity at constant volume for an ideal gas with mean molecular
 weight $m_H/2$; $m_H$ is the mass of a hydrogen atom; $m_*\equiv {M\over M_\odot}$;
 and we assume that $r_*\equiv R_{0}/R_\odot\sim R_{*}/R_\odot$, where $R_{0}$ is  the  radius at the coronal base.
We have used 
 $\dot M=4\pi R_0^2\rho_0 U_0$, where subscripts $0$ indicate values at the coronal base.
 We used  $p_0\sim \rho_0 c_s^2$ for the associated pressure, and  have used the low $r$ limit of Eq. (\ref{2.6}) for $U_0$. 
Plugging in the numerical values for the constants  gives
  \beq
F_{W1}=3.1 \times 10^6 p_0 {\tilde T}^{1/2} e^{3.9{m_*\over r_*}(1-{\tilde T}^{-1})}.
 \label{fwind}
 \eeq

For the x-ray radiation flux we  have
 \begin{eqnarray}
 F_{x}&=&{j_0 p_0^2 \over 4 k_B^2 T^2}\cdot {2k_BT\over m_H g}
 ={j_0 p_0^2 \over 4 k_B T}\cdot {R_*^2\over m_H G M_*} \nonumber \\
 &=&1.24\times 10^6{ p_0^2\over {\tilde T}^{5/3}}{r_*^2\over m_*},
  \label{frad}
 \end{eqnarray}
 where we used   $j_0= 10^{-17.73}T^{-2/3} \rm {erg\cdot cm^3 /s}$ for the radiative loss function, which works well for the 
 range $ 2 \times 10^6 {\rm K} \le T \le 10^7$K  
 \citep{Aschwanden2004}, and not badly as an average extending down to  $4\times 10^5$K. 
 
As in  \cite{Hearn1975}, we assume  that the   fate of  conduction  is heat transport toward the chromosphere where the energy is
 re-radiated  at  a slightly lower temperature.   But we  also include a multiplicative  solid angle correction fraction 
  ${\tilde \Theta}/4\pi \le 1$ 
because conduction down from the corona  is non-negligible only along the 
fraction of the solid angle covered with field lines perpendicular to the surface.
 We then have
 \begin{eqnarray}
 F_{C}
&=& 4.19\times 10^{15} p_0(\kappa_0j_1)^{1/2}T^{3/4}{{\tilde \Theta}\over 4\pi}\nonumber \\ &=& 4.26\times 10^6 p_0{\tilde T}^{3/4}{{\tilde \Theta}\over 4\pi},
 \label{fcond}
 \end{eqnarray}
 where  $j_1\simeq 2 \times 10^{22}{\rm erg\cdot cm^3/s}$  approximates 
 the radiative loss function over the range $ 10^4 {\rm K} \le T \le 10^6$K, 
 and ${\kappa_0=1.1 \times 10^{-6}{\rm erg\over cm \cdot s \cdot K^{7/2}}}$ is the thermal conduction coefficient \citep{Athay1990}.
 
Using Eqs. (\ref{fwind}), (\ref{fcond}), and (\ref{frad}) in Eq (\ref{equ}),  and assuming that $p_0$
is a constant when taking the partial derivatives with respect to $T$, we
then obtain an expression for $p_0$ as a function of ${\tilde T}_0\equiv {T_{0}\over 3\times 10^6K}$, 
where $T_0$ is the coronal temperature  at 
 equilibrium.  The result is
\beq
\begin{array}{r}
p_0=
{m_*\over r_*^2}
{1.6}{\tilde \Theta} {\tilde T}_0^{29\over 12}+ {m_*\over r_*^2}{0.75}{\tilde T}_0^{13\over 6}
e^{3.9{m_*\over r_*}\left(1-{1\over {\tilde T}_0}\right)}\\
+{m_*\over r_*^2}{2.34}{\tilde T}_0^{7\over 6}
e^{3.9{m_*\over r_*}\left(1-{1\over {\tilde T}_0}\right)}.
\end{array}
\label{P0}
\eeq
This  can then be used in Eqs, (\ref{fcond}),  (\ref{frad}), and  (\ref{equ})
to obtain $F_{W1}$, $F_C$, and $F_x$. 
 
For a solar coronal
temperature $T_{x,\odot}\sim 1.5\times 10^6{\rm K}$ \citep{Aschwanden2004},
we find that $F_C$ must be subdominant to account for the solar values ${\dot M}(T_{x,\odot}) \sim  {\dot M}_{\odot} $, and
$ {{\mathcal L}_X({T}_{x,\odot}) \sim {\mathcal L}_{X\odot}}\sim 1$.
This requires  the plausible constraint that  ${\tilde \Theta} \le 1/10$, which also makes
 the conduction negligible for our equilibrium calculation for stars younger than  the sun as we shall see.   We intuit that the value of ${\tilde \Theta}$ is 
strongly correlated with the sunspot covering fraction $\Theta/4\pi\le 0.1$ discussed later (above Eq. {\ref{lx}).
If in fact  $\Theta \sim {\tilde \Theta}$, then ${{\tilde \Theta} \over 4\pi} \le 0.1$ is a self-consistent  upper limit for most of the age range of main-sequence stars in our framework.


Using Eq. (\ref{P0}) in  Eq. (\ref{frad}) and Eq. (\ref{fwind})  along with
  $l_x \equiv {{\mathcal L}_x\over {\mathcal L}_{x\odot}}={F_{x}\over F_{x,\odot}}$, 
  (assuming $m_*=r_*=1$ for all time), and  $\dot m\equiv {\dot M \over \dot M_{\odot}}$,
we obtain 
\beq 
l_x  \simeq
Exp\left[ln({\tilde T}_0)+7.8{m_*\over r_* {\tilde T}_0 }\left({{\tilde T}_0\over {\tilde T}_{0\odot}}-1 
 \right) \right]\simeq {\dot m}.
\label{dotm}
\eeq
As we will focus on coronae for which ${\tilde T} < 2$, 
the logarithm in the exponent can be ignored and the equation can then be conveniently inverted to obtain
\beq 
{\tilde T}_0  \sim [2-{0.26}({Ln}( l_x))]^{-1} = [2-{0.26}( { Ln}({\dot m}))]^{-1}
\label{temple}
\eeq

Fig. \ref{fig1}a shows how the coronal wind, conduction, and radiation fluxes 
 ${\dot m}={F_{W1}{\tilde T}_{0,\odot} \over F_{W1,\odot}{\tilde T}_0}$, ${F_C\over F_{C,\odot}}$, and $l_x={F_x\over F_{x,\odot}}$   
 change as a function of the equilibrium temperature, ${\tilde T}_0$, for ${\tilde \Theta} =1/10$.
In this plot we have normalized the x-axis in units of $ 3\times 10^6$K
and the y-axis quantities in units of their present solar values. As such, 
all of the normalized curves pass through unity at $\tilde T=0.5$.
The plot therefore shows  relative contributions of each quantity compared to their present solar values.   (Unlike Fig \ref{fig1}a,  in which quantities are normalized to their respective solar values, Fig \ref{fig1}b    conveys the  relative contributions of fluxes to each other because each quantity is normalized to the same constant.)

In Fig. \ref{fig1}a-d,  we have used a vertical line to demark  two  regimes defined
by their  x-ray temperature with respect to the solar average value. 
Regime I is defined by  ${\tilde T}<{\tilde T}_\odot =0.5 $, and corresponds to solar-type stars  older than the sun). Regime II is defined by  ${\tilde T}>{\tilde T}_\odot =0.5 $, and corresponds to solar-type stars younger than the sun).  Fig. \ref{fig1}a, 
  shows that  thermal conduction is  fractionally more  important than for the sun  for most all of regime I,   whilst being subdominant  in most all of regime II.   This distinction is  important because conduction could  divert power from the wind in regime I,  in turn suppressing angular momentum loss. This may help explain why the stellar rotation  period-age relation seems to flatten for stars older than the sun\cite{vanSaders2016}.   
Also for most all of regime I,  $\dot m$  is a  much steeper function of temperature than  $l_x$ whereas for regime II, the the two scale  more closely in lock step. 
That is,  $dl_x/dT < d{\dot m}/dT$ in regime I whereas  $dl_x/dT =d{\dot m}/dT$ in regime II.
If averaged over both regime I and regime II therefore, this combination is not too far off from
the observed trend  ${\dot M}\propto F_X^{1.34 \pm 0.18}$   \cite{Wood2005,Wood2014}.  

The purple dashed straight line of Fig. \ref{fig1}a   corresponds to  ${\tilde T}^5$ and the dotted red line is Eq. \ref{temple} expressed as   ${\tilde T} \exp\left[-3.9{m_*\over r_*}\left({1 \over {\tilde T} }- {3\times 10^6{\rm K}\over T_{0,\odot}}\right)
\right]^2$, being a reasonable match to the $l_x\sim {\dot m}$ curve in regime II.
The straight line  $F_x\propto T^5$ (straight line in Fig. \ref{fig1}a) is  not 
too far off from the average slope over the observed range, even where the exponential  fit of Eq. (\ref{dotm}) is best.  
In Fig {\ref{figT_LX} we have also compared  Eq. (\ref{dotm}) directly with the data of \cite{Johnstone2015} for two different values of
the solar coronal equilibrium temperature used therein for solar minimum and maximum
$T_{x,\odot1}=0.97\times 10^6$K  and $T_{x,\odot2}=2.57\times 10^6$K.
Fig \ref{figT_LX} shows that Eq. (\ref{dotm}) does  reasonably well to match the data.
We focus  on regime II in our time-evolution solutions of the subsequent sections of this paper,
leaving study of regime I for future work.

Although the total wind power escaping from the scale height of the corona  is given by  ${\mathcal L}_{W1}\sim {\dot M}c_V T$,  the
 total  wind power escaping from the star is at least ${\mathcal L}_{W} \simeq {1\over 2} {\dot M} v_{esc}^2$.   Assuming constant $\dot M$,  mean molecular mass $m_p/2$,   and thus
$c_s= 2k_bT/m_p$  and $c_V=3k_b/m_p$  \citep{Hearn1975},  the relation between 
the total wind power and the coronal wind flux is
 \beq
 {\mathcal L}_{W} = 4\pi R_o^2  F_{W1}{2v^2_{g} \over 3c_s^2} =4\pi R_o^2  F_{W1}{5.1m_*\over {\tilde T} r_*},
  \label{lwtot}
   \eeq
 where $v_g^2\equiv GM_*/R_0$.
Fig. \ref{fig1}b  shows this total integrated wind power $ {\mathcal L}_{W}$ (blue) 
 along with  ${\mathcal L}_C=4\pi R_o^2  F_C$ (green) , and ${\mathcal L}_x=4\pi R_o^2  F_x$ (orange) with all three quantities normalized to the same constant ${\mathcal L}_{x,\odot}$.
 We see that ${\mathcal L}_C<< {\mathcal L}_x\simeq  {\mathcal L}_{W}$ in regime II.  Note also that when the equilibrium $p_0$ is plugged back into the expressions for $F_x$ and $F_W$, the ratio of ${\mathcal L}_W/{\mathcal L}_x $ scales with $m_*\over r_*$. Therefore,  although we focus on $m_*=r_*=1$, for the temperature range of interest  we would predict ${\mathcal L}_x \sim {\mathcal L}_W$ for a range of late-type stars,  largely independent of mass since $r_*\sim m_*^{0.9}$ for the relevant stellar models 
  \citep{Kippenhahn2012}.

Figs. \ref{fig1}c \& d are similar to  Figs. \ref{fig1}a \& b but apply when the last exponent in Eq.  (\ref{14}) is reduced  by a factor of 2,   highlighting the sensitivity of  the temperature dependence. Here  the power-law fit $l_x\sim {\dot m} \sim {\tilde T}^5$
more accurately matches  regime II than the  exponential of Eq. ({\ref{temple}).

 With the physics included in our present model, we do not identify a 
huge drop  in mass loss rate at high ${\mathcal L}_X$ as suggested 
by 2 data points in  \cite{Wood2014}. 
More data are needed to confirm this is a statistically significant effect.
There does not seem to be evidence  a corresponding change of magnetic configuration 
\citep{Vidotto2016}, so that would not  provide an explanation if the effect were real.
  \cite{Suzuki2013} suggested that a decrease in mass
  loss for young active stars might arise because the increase in coronal density 
  could lead to runaway cooling that drains the thermal energy 
  that would otherwise  accelerate the wind. 
  This model does not employ  coronal equilibrium arguments that we have used herein
  to determine our coronal properties. We leave  further consideration of this
  for future work.

 \begin{figure}
 \centering
{\includegraphics[width=0.77\columnwidth]{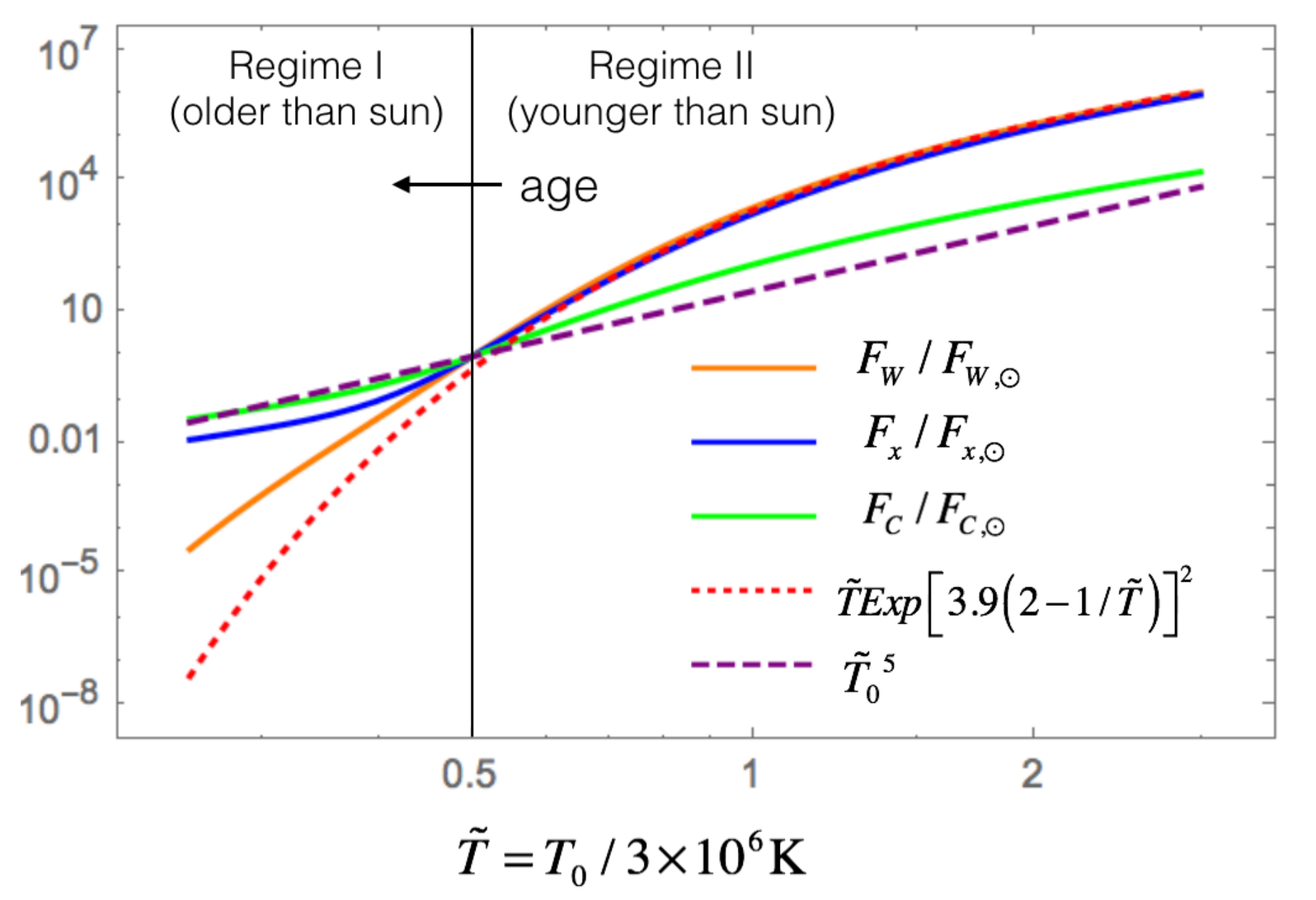}}
{\includegraphics[width=0.75\columnwidth]{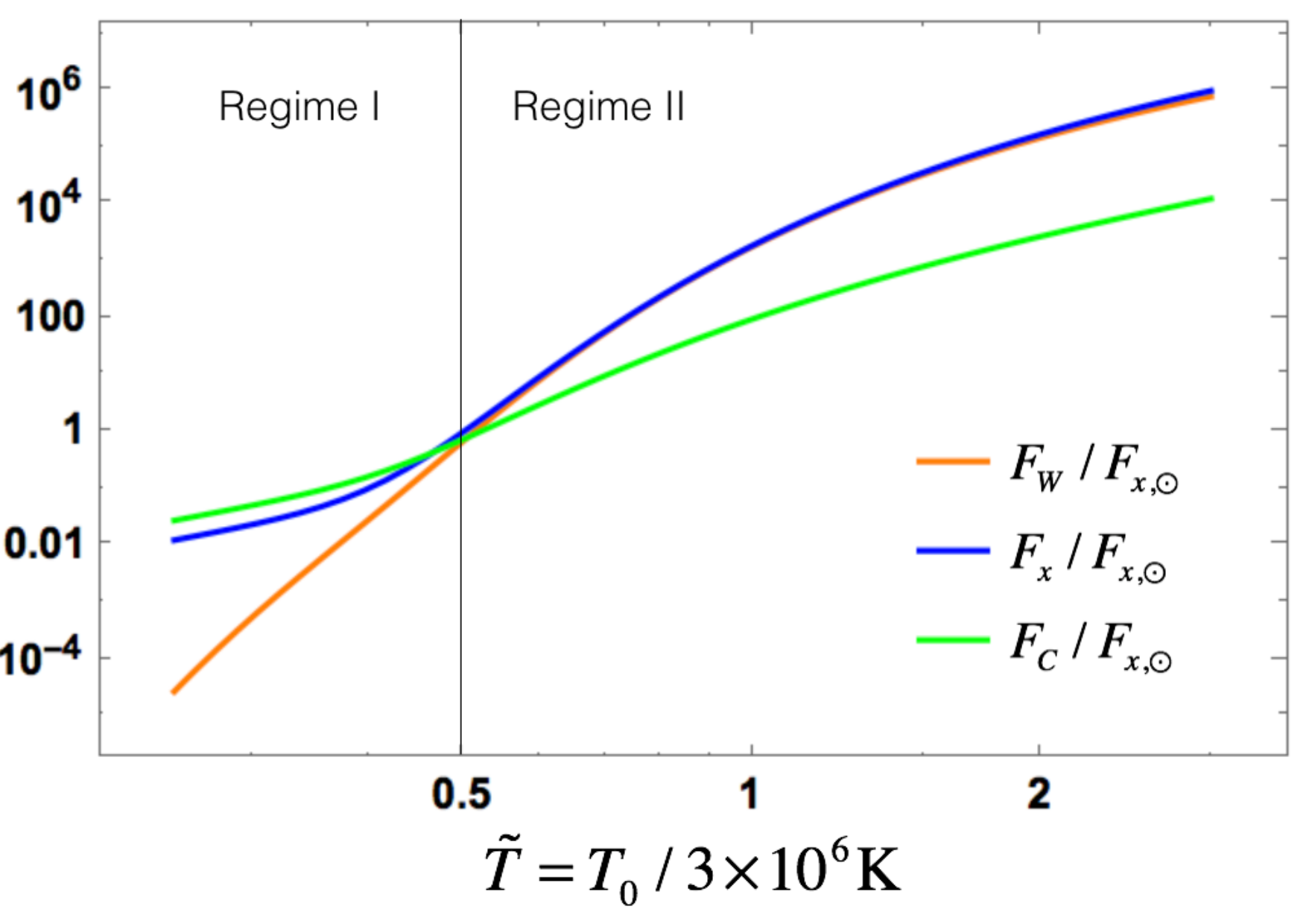}}
{\includegraphics[width=0.77\columnwidth]{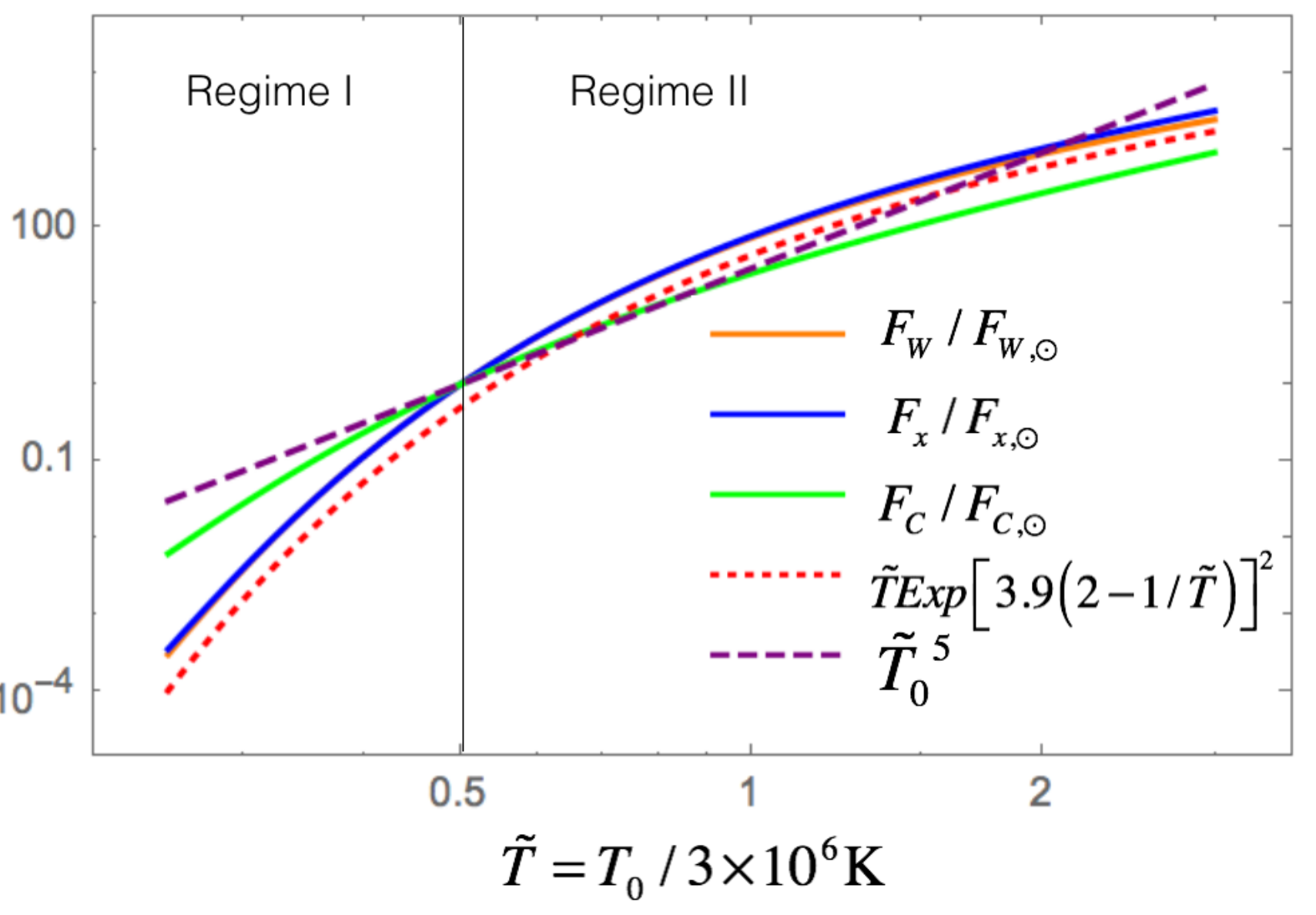}}
{\includegraphics[width=0.77\columnwidth]{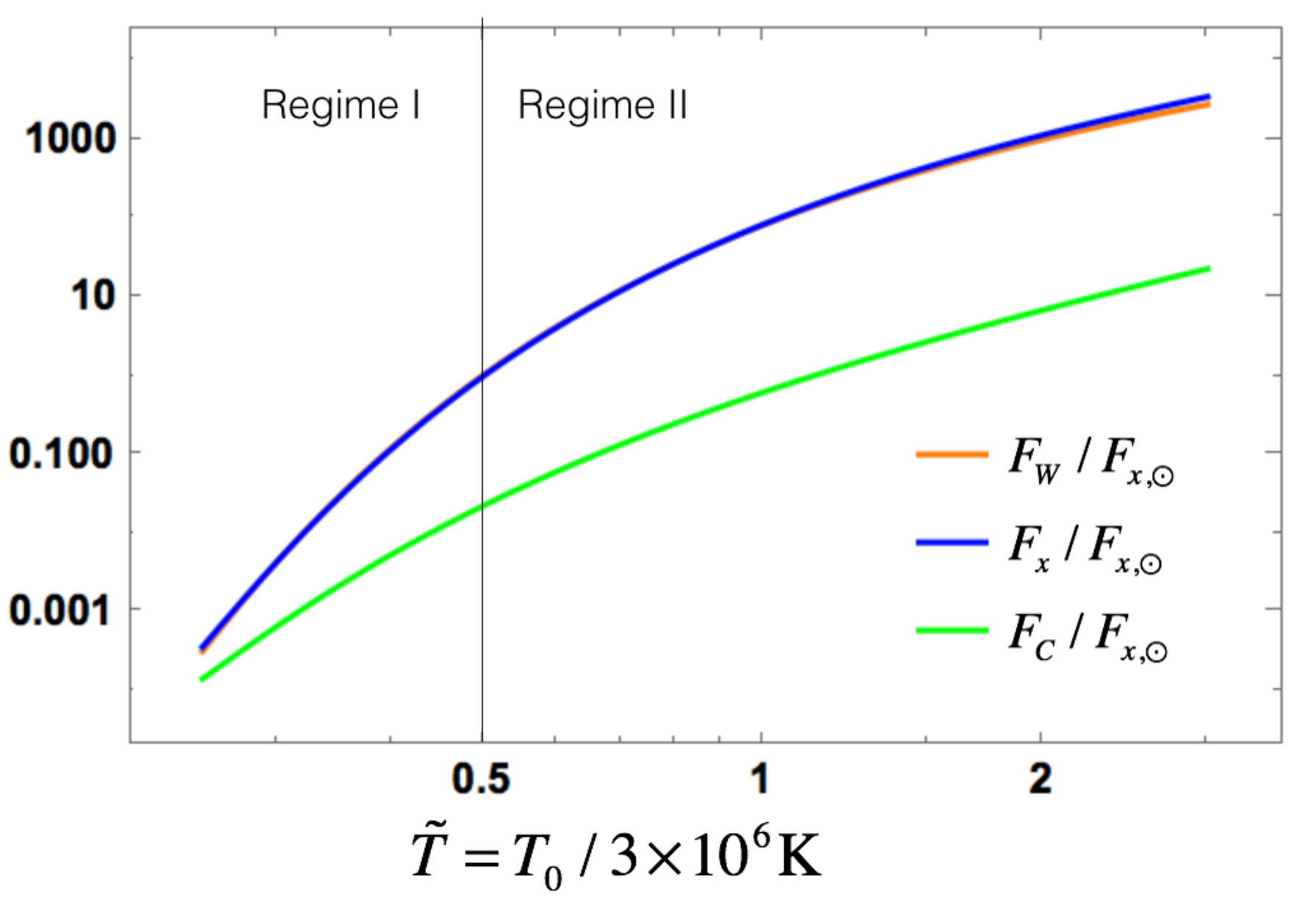}}
 \caption{
 Panel $a$ shows $F_{W1}/F_{W1,\odot}$ (blue); $F_C/F_{C,\odot}$ (green);  $l_x=F_x/F_{x,\odot}$ (orange);  ${\tilde T}^5$ (purple dashed);  $ {\tilde T}Exp\left[-3.9\left({1 \over {\tilde T} }- {3\times 10^6{\rm K}\over T_{0,\odot}}\right)
\right]^2$ (red dotted). The y-axis is in units of  the  solar value for each quantity. 
Regimes I and II  are defined by their x-ray temperature with respect to the solar average value, namely,  ${\tilde T} <0.5$ and  ${\tilde T} >0.5$ respectively for ${\tilde T}_\odot=0.5$.
Panel $b$ shows  ${\mathcal L}_X$, ${\mathcal L}_W$, and ${\mathcal L}_C$ all normalized to  ${\mathcal L}_{X\odot}$
showing that  coronal equilibrium for Regime II  leads to ${\mathcal L}_X\simeq {\mathcal L}_W$..  Panels a and b show that ${\mathcal L}_C$
is subdominant for most all of regime II but dominant in most of  regime I where the wind
is correspondingly subdominant.  Panels c and d have the same information as panels a and b but 
for a case in which the last exponent of   
Eqn. (16) is reduced  by a factor of $2$, showing that  the power-law fit $l_x\sim {\dot m} \sim {\tilde T}^5$ is not  a bad approximation to $\dot m$ and $l_x$ for $0.5\le {\tilde T}\le 2$. In all panels we have used $\tilde \Theta =0.1$ (see text).}
 \label{fig1}
\end{figure}

\begin{figure}
\centering
{\includegraphics[width=0.9\columnwidth]{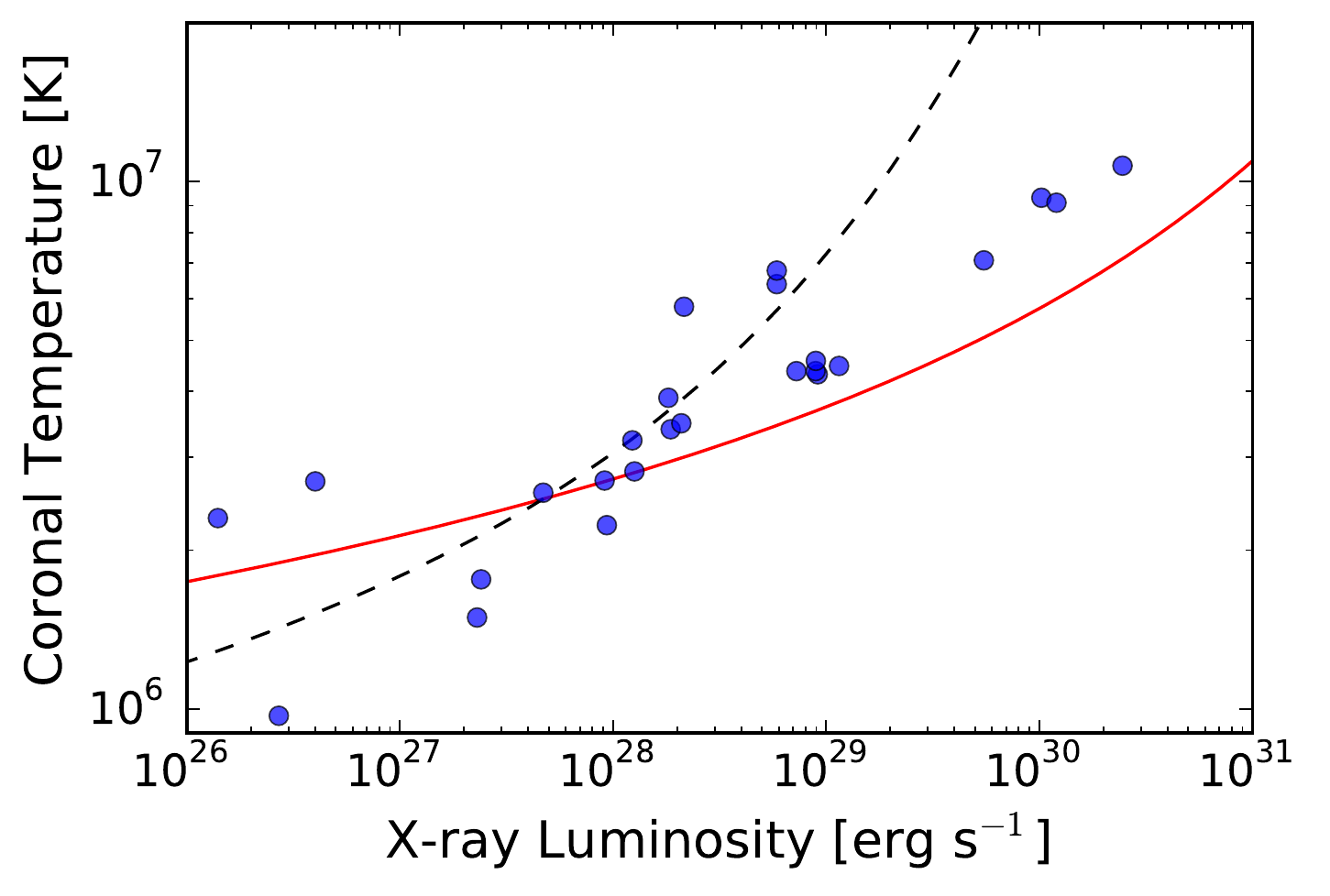}}
 \caption{Comparison of coronal equilibrium  temperature $T_0$  vs. x-ray luminosity from 
Eq.\ref{dotm}  compared to data  (Johnstone et al. 2015).
The solid and dashed curves respectively correspond  to the equilibrium temperatures  
 for solar minimum and maximum,  $T_{x,\odot1}=0.97\times 10^6$K  and $T_{x,\odot2}=2.57\times 10^6$K (Johnstone et al. 2015).  The plot shows that Eq. (\ref{dotm}) is consistent with  the data.}
 \label{figT_LX}
\end{figure}
 \subsection{ X-ray luminosity as function of  magnetic field strength} 
 
Having established that  $l_x \simeq  {\dot m}$ in regime II, we now connect both of these quantities
to the source of magnetic energy.  \cite{Blackman2015}  estimated 
 $\mathcal{L}_{X}$  from the dynamo-produced magnetic energy rising up through the convection zone. 
 To accommodate our present result that
that   1/2 of  source magnetic energy goes into  $\mathcal{L}_{X}$, the relevant expression
becomes
\beq
\begin{array}{r}
\mathcal{L}_{X}\simeq 0.5  \mathcal{L}_{mag} \simeq  0.5{B^2 u_b\over 8 \pi} \Theta r_c^2
= {1\over 3}\left({B^2 \over 8 \pi}\right)^2{ \Theta r_c^2\over \rho v},
\end{array}
\label{lmagsimple1}
\eeq
  where $B\sim {B}_\phi$ and $\Theta/4\pi$  is the solid angle 
  fraction through which the field rises (proportional to, if not   equal to  $\tilde \Theta$ of the previous section), and we have used the expression for $u_b$   below Eq. (\ref{16}).
Using  Eq. (\ref{bsat}) in Eq. (\ref{lmagsimple1})
then gives
 \beq
\begin{array}{r}
{\mathcal{L}_{x} \over \mathcal{L}_* }\simeq
 \left({L_\alpha\over r_c}\right)^{2/3}\left({s^{1/3}\over 1+ s{\tilde Ro}}\right)^{2} \Theta \left({3\pi\over 8 }\right)^{1\over 3}\left({q_\alpha cos \theta_s \over 6}\right)^{2/3},
\end{array}
\label{lmagsimple3}
\eeq
where   we have invoked
 $
\mathcal{L}_*
  \simeq   4\pi r_c^2\rho {v^3},
$
 the luminosity  associated with the convective heat flux 
through the convection zone \citep{Shu1992}.
For the sun ${\tilde Ro}\sim 2$ and $L_\alpha \sim 2r_c/5$.

We posit that $\Theta/4\pi \sim a_{spt}$ 
 the  areal fraction through which  the strongest buoyant fields   penetrate, approximately equal  to  the areal fraction of sunspots.
For the sun  $a_{spt,\odot} \lsim 0.005$   \citep{Solanki2004}.  
We allow $\Theta$ to depend on $\mathcal{L}_X/\mathcal{L}_*$ and take 
$\Theta=\Theta_\odot[({\mathcal L}_X/ {\mathcal L}_{\odot})/ ({\mathcal L}_{X\odot}/ {\mathcal L_
\odot})]^{\lambda}$ \citep{Blackman2015}.
For $\lambda = 1/3$, this implies that a factor of $10^3$ increase in the x-ray to bolometric luminosity ratio would 
imply  $\Theta /4\pi =10a_{set,\odot}  < 0.1$, making self-consistent our assumption that ${\tilde \Theta} \le 0.1$ above Eq. (\ref{fcond})
if  $\Theta\sim {\tilde \Theta}$.
 Combining with Eq. (\ref{lmagsimple3})  we then  have for ${\mathcal L}_*= { \mathcal L}_\odot$
\begin{eqnarray}
l_x &\equiv& {1\over 1.4-0.4t}\left({s \over s_\odot }\right)^{2\over 3(1-\lambda)}
\left({{1+s_\odot  {\tilde Ro}} \over  {1+s {\tilde Ro}}}\right)^{2\over 1-\lambda}\nonumber \\&=
&b_r^{4\over 1-\lambda},
\label{lx}
\end{eqnarray}
where the factor $(1.4-0.4t )^{-1}$ approximates the increase  in  sun-like bolometric luminosity  with
  time $t$ in unit of solar age   \citep{Gough1981}, and 
 the latter equality of Eq. (\ref{lx}) follows from Eq. (\ref{bsat}) of \cite{Blackman2015}
who showed that $0\le \lambda\le 1/3$ corresponds to $2\le \zeta\le 3$ (in Eq \ref{rossby1}) in the unsaturated
rotator regime.  Increasing $\lambda$ increases the saturated $l_x$.
Eq. (\ref{lx}) implies that $g_L(t)=(1.4-0.4t)^{\lambda-1\over 4}$ in  Eq. (\ref{2.42}).

\section{Angular Momentum Evolution}
In the previous sections, we have made theoretical progress by coupling the magnetic field strength, x-ray emission and mass-loss. To  finally compute the global time-evolution of these quantities we need to know how our mass-loss rates and magnetic field strengths give rise to a torque on the star and spin it down.  For simplicity we adopt a simple analytic formalism, although coupling with more detailed MHD simulations should be possible in the future. 

Physically, we  consider the Parker spherical wind solution to  propagate along  radial large scale fields out to the Alfv\'en radius where the toroidal field and radial field start to become comparable in the equatorial plane. We assume that the thermal energy driving the Parker wind is sourced via magnetic dissipation such that  magnetic energy  need  not appear explicitly in the radial momentum equation for the wind.
We correspondingly  assume that the large scale Poynting flux does not contribute significantly to the wind acceleration. 
This is reasonable as long as the speed at the sonic point computed from the Parker wins is larger than the 
Michel velocity \citep{Lamers1999}, which is self-consistently satisfied for the full range of our present solutions. 
We   consider the angular momentum loss to be determined by that from the equatorial plane \citep{Weber1967}.
  
\subsection{Angular momentum and toroidal field}
Over a time scale  shorter than the stellar spin-down time, the quasi-steady angular momentum
equation of the wind  in the equatorial plane  reduces to \citep{Weber1967,Lamers1999}
\beq
\partial_{r}(r U_\phi) = {B_r r^2 \over {\dot M}}\partial_{r}(r^2 B_\phi),
\label{2.8}
\eeq
where $U_{\phi}(r)$ is the azimuthal wind speed.
Volume integrating the  constraint that
$\nabla\cdot \bfB=0$, implies 
\beq
B_r= B_{r\odot}(r_\odot/r)^2,
\label{2.9}
\eeq
and 
 in combination with Eq. (\ref{2.8})  implies the radial constancy of 
\beq
\mA=r\left(U_\phi-{r^2B_rB_\phi\over {\dot M}}\right)=r^2_*\left(\Omega_*-{r_*B_{r*}B_{\phi*}\over {\dot M}}\right),
\label{2.10}
\eeq
where the last equation follows because $\mA$ can be computed at any wind radius, including the base.

The radial Alfv\'en speed is
\beq
u_A\equiv{B_r \over \sqrt {4\pi \rho}} 
\label{2.11}
\eeq
and 
 the Alfv\'en Mach number 
\beq
M_A\equiv U_r/u_A,
\label{2.12}
\eeq
where  $U_r(r_A)=u_A  $  defines the Alfv\'en radius  $r_A$.
The steady-state (on time scales short compared to the spin-down time)
induction equation for the electric field $\bf E$ is 
 $\curl \bfE=0$  and in spherical coordinates for the equatorial plane implies 
\beq
B_\phi = (-\Omega_* B_r +U_\phi B_r)/U_r,
\label{bphi}
\eeq
Using Eqs.  (\ref{2.11}),  (\ref{2.12}), (\ref{bphi}), and 
$\dot M=4\pi r^2\rho U_r$  in  the first equality of (\ref{2.10}) and solving for $U_\phi$ gives (e.g. \cite{Lamers1999})
\beq
U_\phi={\Omega r}\left({\mA M_A^2\over r^2\Omega_*}-1 \over { M_A^2-1}\right).
\label{2.16}
\eeq
so that a finite $U_\phi$ at $r=r_A$ implies  
\beq
\mA=r_A^2 \Omega_*.
\label{L}
\eeq
Plugging Eq. (\ref{L}) back in to Eq. (\ref{2.10}) gives 
\beq
{r_A\over r_*}=\left(1-{r_*B_{r*}B_{\phi *}\over \dot M\Omega_*}\right)^{1/2}
\label{2.23a}.
\eeq

  Eq. (\ref{2.23a}) 
depends on  $B_{\phi*}$, the only quantity on the right side for which we do not
so far have a scaling in terms of $\Omega_*$ or present day solar values.
We need a separate equation for ${r_* \over r_A}$ that is independent of $B_{\phi*}$ in order to  also obtain  independent equations for both $B_{\phi*}$ and $r_A/r_*$ as a function of the
present  solar values.
To obtain the needed equation we note that Eqs. (\ref{2.9}), (\ref{2.11}) and (\ref{mdotdef})
 imply
\beq
{r_A^2\over r^2_*}={B_{r*} \over B_{r,A}}={B_{r*} \over(4\pi \rho(r_A))^{1/2}u_A}={B_{r*} r_A\over {\dot M}^{1/2} u_A^{1/2}},
\label{2.18}
\eeq
or
\beq
{r_A\over  r_*}={B_{r*} r_*\over {\dot M}^{1/2} u_A^{1/2}}={b_{r*} \over {\dot m}^{1/2} {\tilde u}_A^{1/2}} {B_{r\odot} r_*\over {\dot M}_{\odot}^{1/2} u_{A\odot}^{1/2}},
\label{2.19}\eeq
where ${\tilde u}_A\equiv {u_A\over u_{A\odot}}$ and
we have used the definitions of $\dot m$ and $b_r$ from Eqs. (\ref{2.42}) and (\ref{dotm}).
Rearranging Eq. (\ref{2.23a}) and using  Eq. (\ref{2.19})  then gives
\begin{eqnarray}
b_{\phi*}&\equiv& {B_{\phi*}\over B_{\phi\odot}}\nonumber \\ &=&-{{\dot M} \Omega_* \over r_* B_{r*}B_{\phi\odot}}\left[ {r_A^2\over r_*^2}-1\right]\nonumber \\
&=&-{{\dot m} \omega_* \over b_{r*}}{{\dot M}_\odot \Omega_\odot \over  r_* B_{r\odot}B_{\phi\odot}}\left[ {r_A^2\over r_*^2}-1\right],
\label{2.17}
\end{eqnarray}
where $\omega_*\equiv \Omega/\Omega_\odot$. 
The needed expression for  $u(r_A)=u_A$ for use in  Eq. (\ref{2.19}) 
(and thus in Eq. (\ref{2.17})) 
can be approximated from the two asymptotic forms 
 Eqs. (\ref{2.6}) and  (\ref{2.7}) taken at $r=r_A$.
 These latter two expressions then normalized to $u_{A,\odot}$ (the value for the present day sun)
 can then be combined in the form
\small
\begin{eqnarray}
&&{\tilde u}_A(t)\equiv u_A/u_{A\odot}=
\sqrt{T\over T_\odot}\times\label{ua} \\
&&\!\!\!\!\!\!\!\!\!\!\!\max\left\{\exp \left[2\left({r_{s\odot}\over r_{A\odot}}\right )\left(1-{T_\odot \over T}{r_{A\odot}\over r_A}\right)\right];
\sqrt{
{Ln \left[{T\over T_{\odot}}{r_A\over r_{*}}{r_{*}\over r_{s\odot}}\right]
 \over   Ln\left[{r_{A\odot}\over r_{s\odot}}\right] }}\right\}\nonumber.
\end{eqnarray}
In practice,  the sonic radius is always below the Aflv\'en radius for our purposes so the latter of the
two approximations actually suffices for present purposes.  
\normalsize

\subsection{Time-evolution of   angular velocity}

Over time scales  $>> r_s/U_r$, the  star will spin-down by 
 angular momentum loss into the wind via  the magnetic field. From  conservation of angular momentum, the evolution of the stellar angular velocity is then
\beq
{\dot \Omega} =- { q\over 0.059 Mr_*^2}{ \mA{\dot M}} =-{{q\dot M}\Omega_* \over0.059 M}{r_A^2\over r_*^2}=-{q\Omega_* \over 0.059 M}{B_r^2 r^2_*\over  u_A}
\label{2.23},
\eeq
where ${{\mathcal I}_\odot \over M_\odot r_\odot^2}=0.059$, ${\mathcal I}_\odot$ is the moment of inertia of the sun, and the inertial parameter
$q>1$ is a numerical factor to reduce this by an amount that depends  
on internal angular momentum transport and the fraction of the star to which the field is anchored.
We  have used Eq. (\ref{L}) and Eq. (\ref{2.19}) for the second and third equalities in Eq. (\ref{2.23})  respectively.  
(We note that the form of Eq (\ref{2.23}) can also be recovered from Eq. 9 of \cite{Matt2012}, by setting our  $q=K_1^2$ therein and taking $f^2 << K_2^2$ and
$m=1/2$ in that equation).
Scaling to  present  solar values, 
Eq. (\ref{2.23})  in  dimensionless form becomes
\beq
{d \omega_* \over d \tau} \equiv -{\omega_*}{qb_r^2 \over m {\tilde u}_A}{B_{r\odot}^2 \Omega_\odot \tau_{\odot}\over  M_\odot u_{A\odot}},
\label{omegadot}
\eeq
where  $\tau_\odot$ is the present day solar age.




\begin{figure}
\centering
{\includegraphics[width=.9\columnwidth]{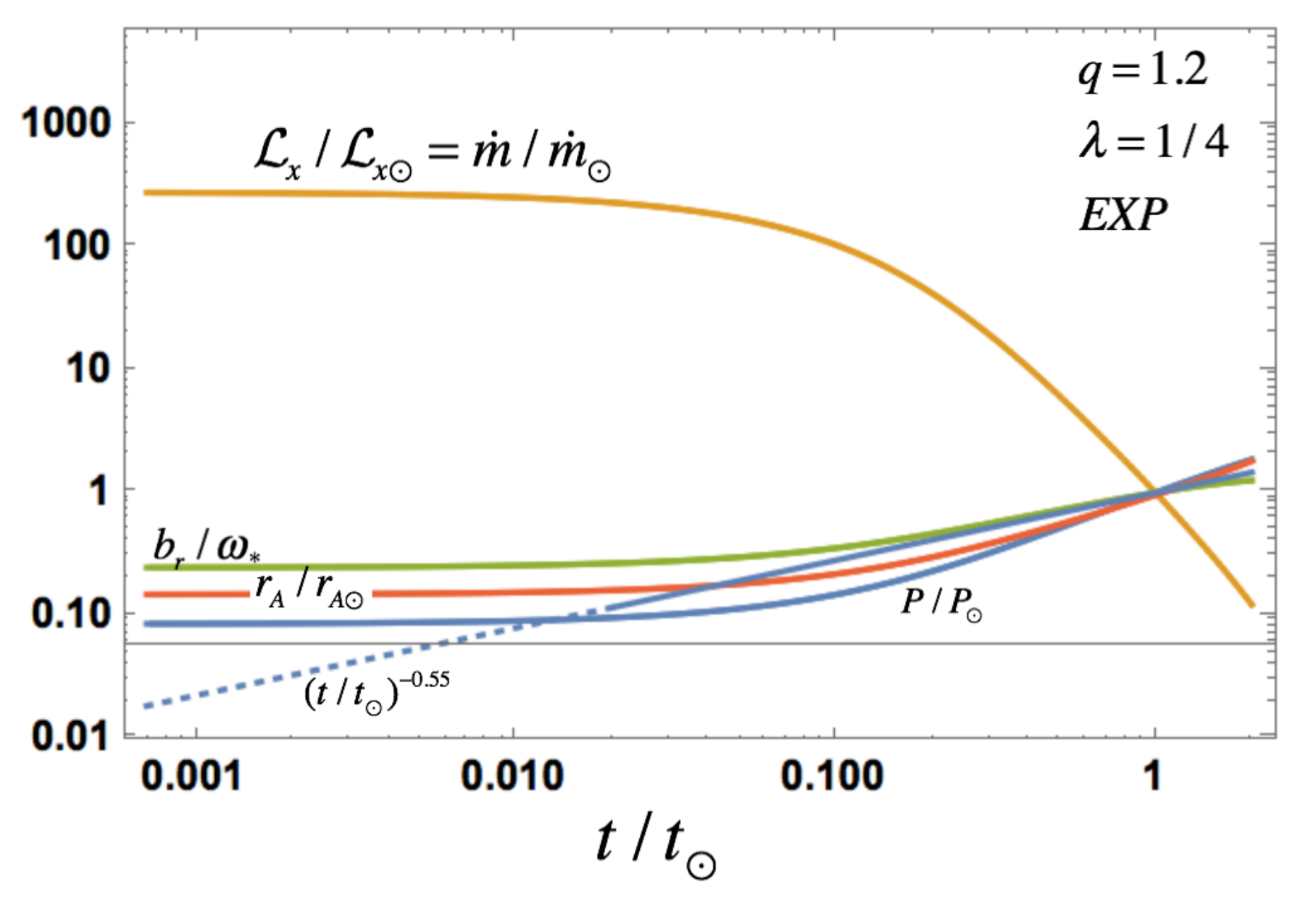}}
{\includegraphics[width=0.9\columnwidth]{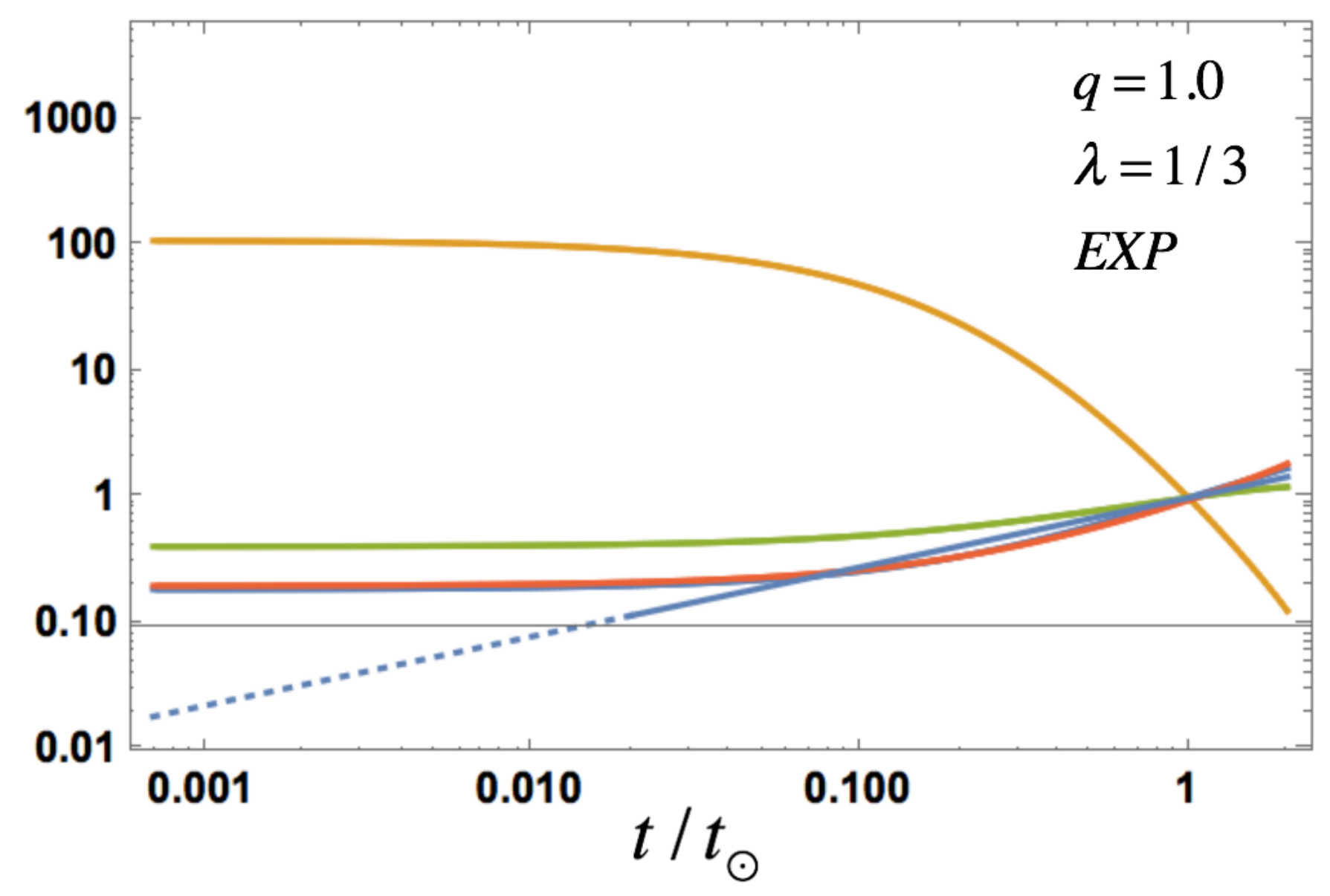}}
{\includegraphics[width=0.9\columnwidth]{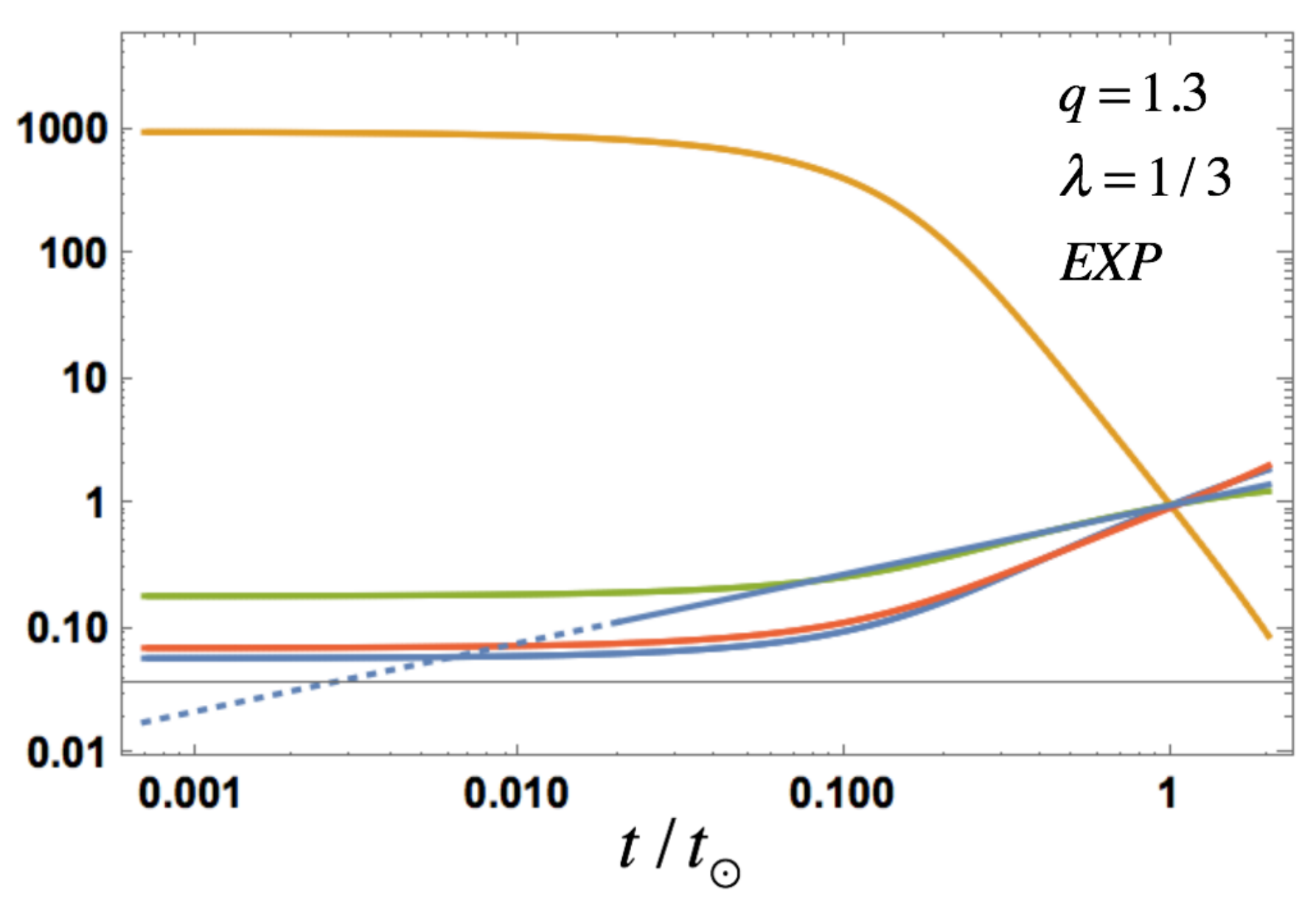}}
{\includegraphics[width=0.9\columnwidth]{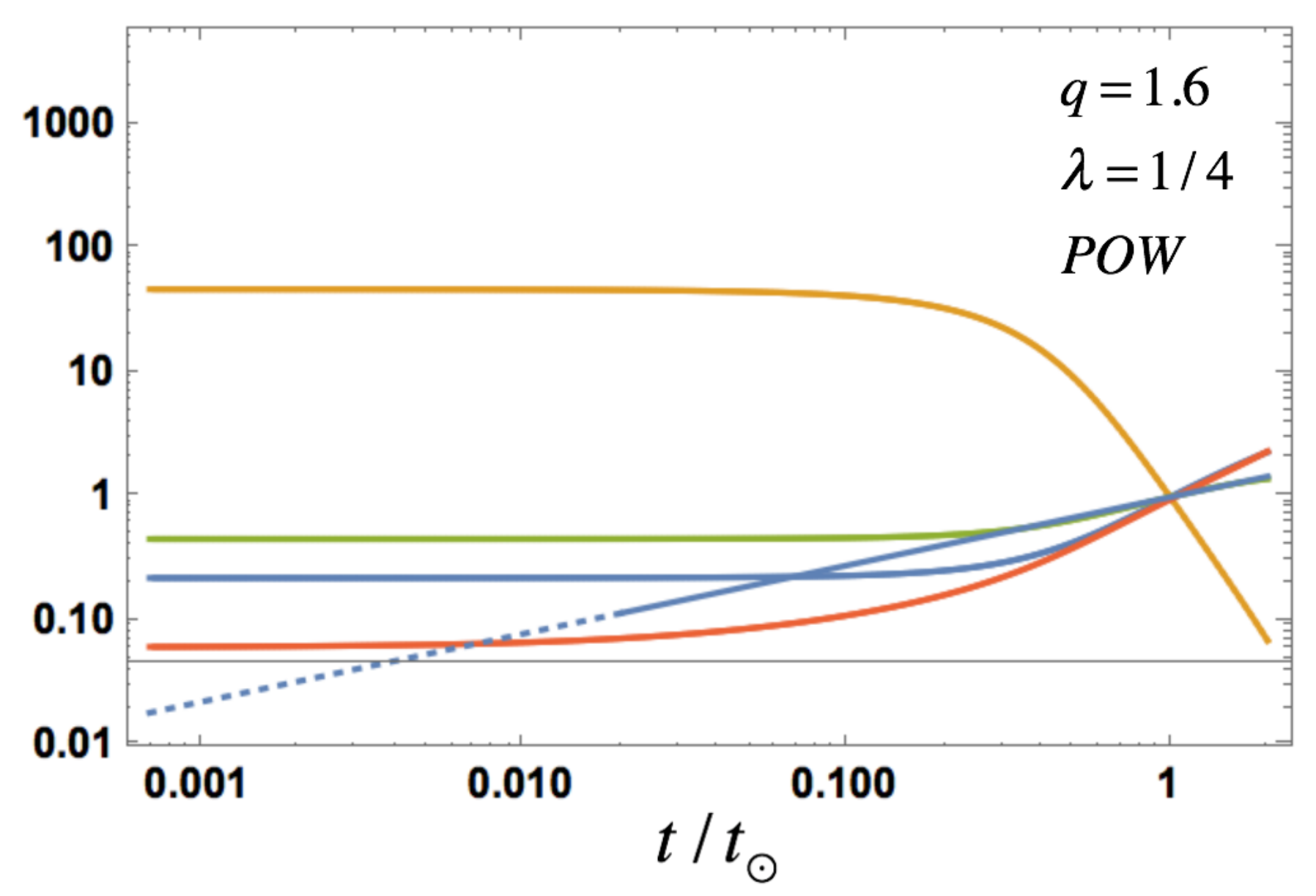}}
 \caption{The   panels (a,b,c,d) each show plots of  $\mathcal{L}_{X}/\mathcal{L}_{\odot}\sim \dot m$ (orange), $b_r/\omega_*$ (green), 
 $r_A/r_{A\odot}$ (red), and  $P/P_\odot$(blue curved)
  compared with the Skumanich law (blue straight) for different $q$ and $\lambda$.
  The latter curve has been extended as a dotted line in the region where it is no longer
  applicable to observations.
  All panels have same colour code. Panels abc corresponds to solutions were
  Eqn (21) is used to relate $l_x$ and $T$
whereas panel  d applies when the last exponent in Eqn (\ref{14}) is reduced  by a factor of $2$  and  the power-law approximation $l_x\sim {\dot m} \sim {\tilde T}^5$
is used. 
  }
 \label{fig2}
\end{figure}
\begin{figure}
\centering
{\includegraphics[width=0.9\columnwidth]{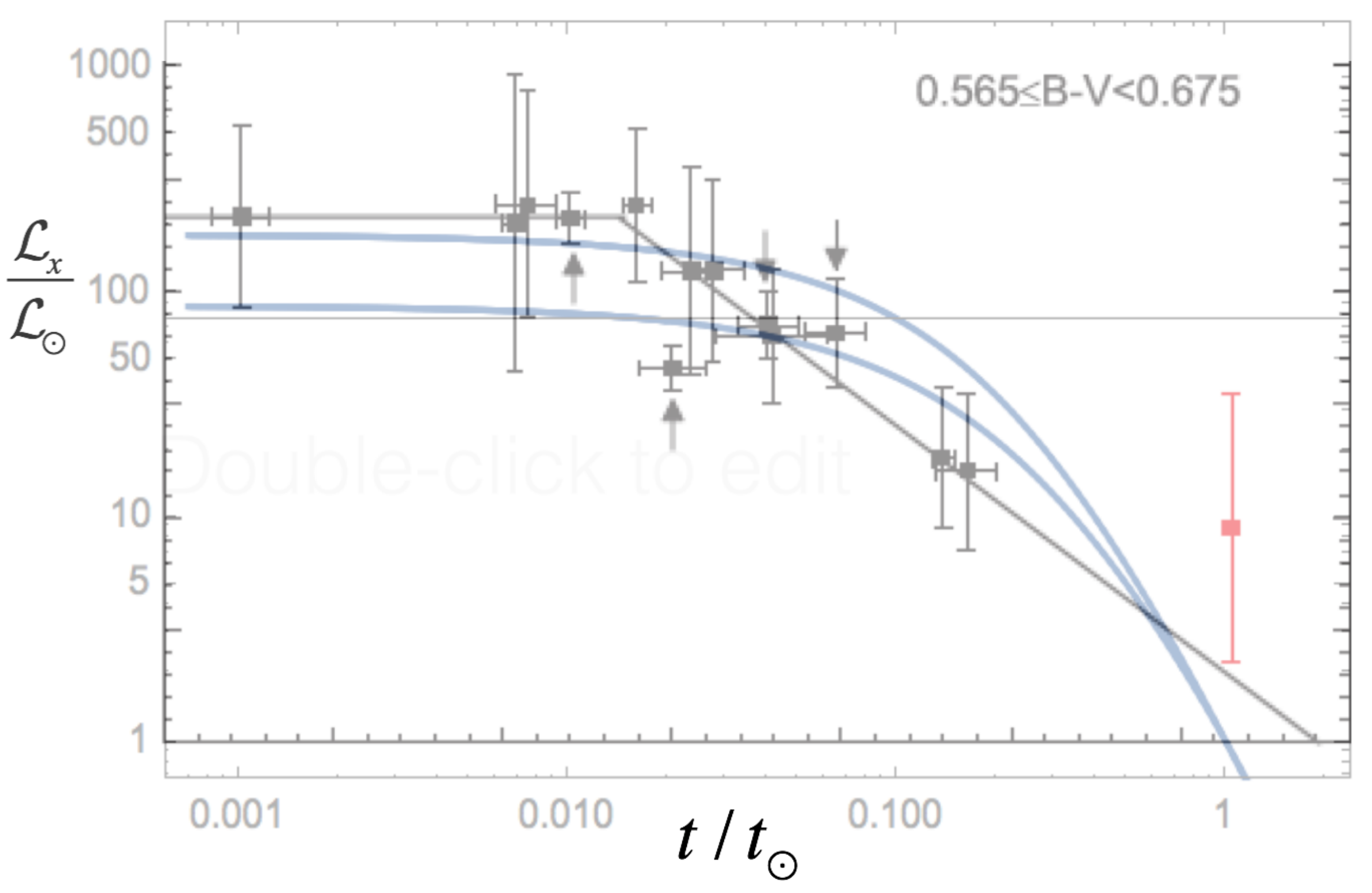}}
{\includegraphics[width=0.9\columnwidth]{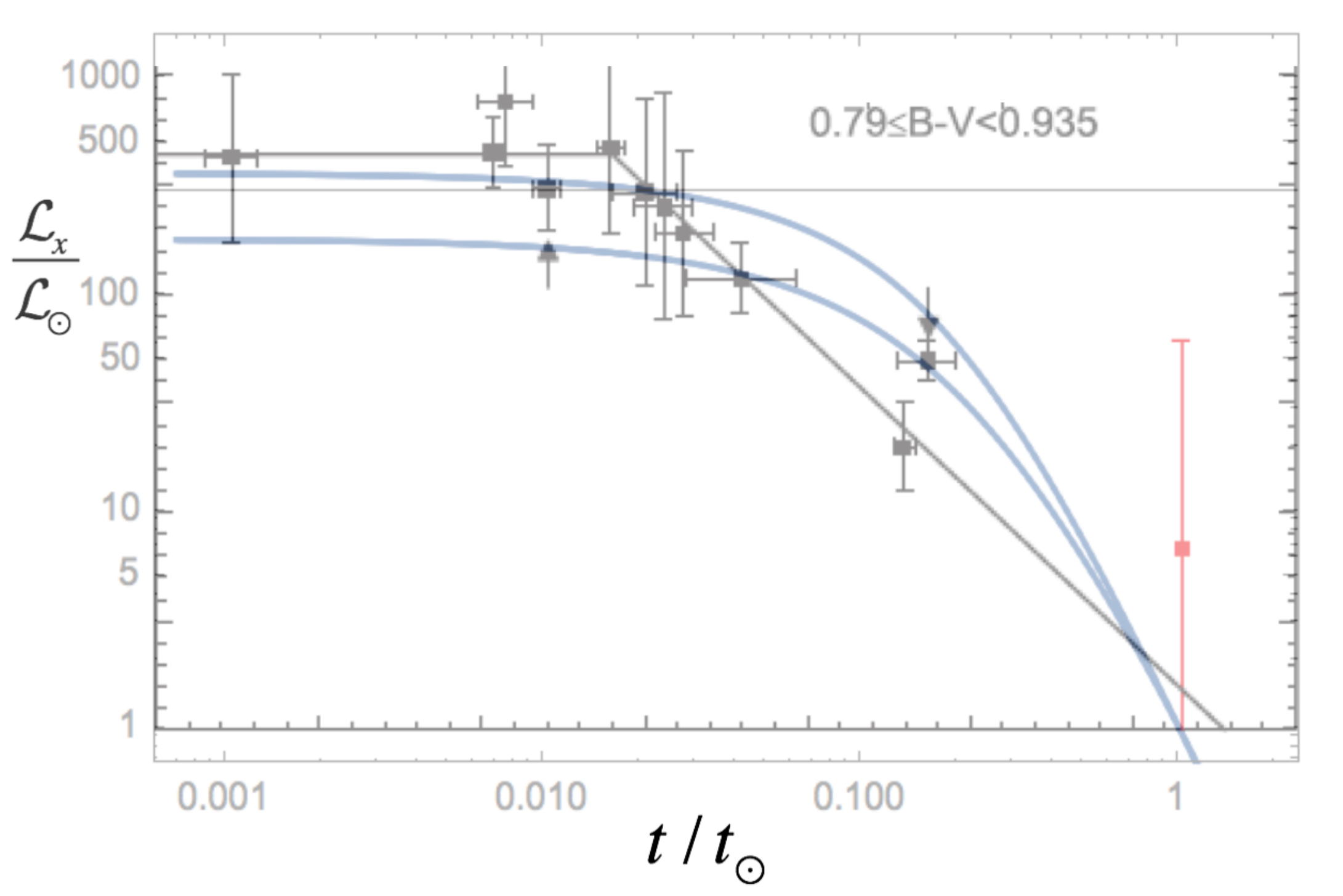}}
 \caption{
    The  two panels show plots of  ${\mathcal L}_x/{\mathcal L}_{bol}$ vs. time (blue lines)
   overplayed with plots from Jackson et al. (2012) who fit the black lines  to 
   the black data points from open clusters selected by color index range (shown on the plot).  The red point is a reference field star not used in their fits.  The top and bottom curves of the top panel  correspond to $l_x$ for $q=1.0$ and $1.1$ and in the bottom panel to  $q=1.1$ and $q=1.2$ respectively.
    Because our solutions are for fixed mass equal to $M_\odot$ but the color of stars of given mass can change with age,
    there is degeneracy a single  star would not evolve in a plot for a singe  color index range.
 The   solar colour index decreases with age and so we simply took  
 plots from the  color index ranges appropriate for the present  sun  ($B-V=0.69$  early sun  ($B-V=0.83$) to
 show examples of how our solutions mesh with data. 
All curves use $\lambda=1/3$.
  }
 \label{fig3}
\end{figure}

\section{Time-Dependent Solutions}

Focusing on the time evolution of a star that  emerges with the current solar 
properties at the sun's age,   we assume that the radius, mass, and thermal convection time of the star do not evolve from the early main sequence to the present time.
Eqs. (\ref{omegadot}),
  (\ref{lx}),  (\ref{dotm}),
  and (\ref{2.17}), along with (\ref{2.19}) and (\ref{ua}) and the present day solar properties then
form a complete set of equations that can  be solved for the coupled time evolution of 
 $\omega_*,  b_r,  l_x\sim  {\dot m}$,  $b_\phi$, and $u_A$.

 \subsection{Basic Properties of the Holistic Solutions}
 
The free parameters of the model are essentially $s$, $q$ and $\lambda$ (with the
measured solar properties as boundary conditions).
We fix the shear parameter $s$ at $s=8.3$ as in \cite{Blackman2015}  because this
value makes the transition from saturated to unsaturated regime  at  $s {\tilde Ro}=1$ or $1/s  =0.12$ consistent with  best fit to observations \citep{Wright2011}.  However the transition is smooth and the
asymptotic regimes ${\tilde Ro} >> 1$, and $ {\tilde Ro}<<1$ are not so sensitive to it.
It depends on the physics of internal differential rotation.   We focus instead on  $q$ and $\lambda$.

  Fig. \ref{fig2}a-c show the time  evolution (in units of solar age)  of $ l_x\sim {\dot m}, 1/\omega_*, r_A/r_{A\odot},$  and $b_r/\Omega$ normalized  to the present solar values   for   values of the inertial parameter $q$  and  $\lambda $ as shown.  
  In our simple model  we treat   $q$ as a free parameter, but note that 
 physical  values require $q>1$ and the  solutions are quite sensitive to the specific choice.   As $q$ is increased, the lower effective momentum of inertia leads to a faster spin-down and thus the early time $l_x$ is higher than for lower $q$ to evolve to the same solar values at the sun's age.   The physically  reasonable range of  $1\le q\le 2$ works best  as we discuss  below with respect to Fig \ref{fig2}.
 In reality,  $q$   is likely a function of time  and would evolve if the  internal rotational coupling of core and envelope evolves. We have assumed that $q$ is a constant for simplicity.
 Fig \ref{fig2}d shows the same information as the previous panels but for a case associated with 
 Figs. \ref{fig1}c \& d in which the exponent in Eqn (\ref{14}) is reduced by a factor of 2.

 The plots of Fig. \ref{fig2}  also show  the  modified 
Skumanich law
\citep{Mamajek2014}
 $1/\Omega = t^{0.55}$ which is known to fit the data well in the unsaturated regime, and can be compared with our dynamical solution.
The Skumanich law does not  apply to stars much younger than  $5\times 10^7$yr
\citep{Gallet2013}
but we show its extension there as well, highlighting the fact that our solutions give 
longer periods at earlier times, perhaps consistent with the "slow rotator" cases of
\cite{Gallet2013}.  The physics of our model is insufficient  for stars with dynamically influential disks,  which means inapplicable for ages  below $~5 \times 10^6$yr. 
 In general,  as $q$ is increased above unity for  fixed $\lambda$  the concavity of our solutions compared to the Skumanich law increases at late times and the intersection between the two curves shifts to small
 times. 

Finally, we note that the purple
 curves of Fig \ref{fig2}. show that $b_r/\omega_*$ remains within an order of unity  solar-like star for our solutions over for at least 3 orders of magnitude in $\tau$
 and remains more constant than the blue curves over both the full time range.
 This results because for the unsaturated regime ( ${\tilde Ro}>> 0.13)$, $B\propto \Omega^{1/2}$ from Eqn.  (\ref{bsat})  
 This is in general agreement with the \cite{Vidotto2014} result 
 mentioned in Sec. 1,  that the field and angular velocity scale similarly with age.
 
\subsection{Example Comparisons of $l_x$ vs. Age with data }
Fig \ref{fig3}.
shows the $l_x$ solutions plotted in same form as Fig. \ref{fig2} for two values of $q$ in each panel overlayed with  data plots of \citep{Jackson2012} for luminosity and age binned by  $B-V$ colour index.   The figure illustrates of how the solutions can connect to the data and what is possible
for further work.   Even though colour might be intended to select a mass range \citep{Pizzolato2003},  the evolution of colour with age
implies that a single star of fixed mass would not be pinpointed  in age using a fixed colour index.
In the figure, we compare our fixed mass ($=M_\odot$) solutions for 
for the parameters indicated with two photometric    classes.  The two classes chosen correspond to the color range associated with the present sun ($B-V= 0.69$) and the color range corresponding to the early sun ($B-V=0.83$) 
 In the two panels we show solutions for $q=1.1$ (top) and $q=1.2$ bottom, 
for the two  $B-V$ photometric bins.  
More detailed modeling which includes solutions for a range of masses and color variations for a given is desired to match theory with data from stellar populations.

Fig \ref{fig3}  shows that the general trends of ${\mathcal L}_{X}/{\mathcal L}_{bol}$ are broadly captured by our model for reasonable ranges of $q$.    However,   if  $q$ is fixed such that the solution fits the saturated early time regime of $l_x$, then the curves
tend to overshoot  in the unsaturated late-time regime. Similarly lowering $q$ to capture the
late time regime of the population then undershoots $l_x$ in the saturated regime.     A better model could include $dq/dt<0$ so that more of the star is coupled to its angular momentum loss as the star ages. In addition, more detailed spin-down torque modeling that produces a different power of $B_r$ in  Eq. (\ref{2.23}) (e.g. \cite{Matt2015}) can modify the turnover  locus and curve shape.
The next subsection  shows that small variations in the average equilibrium temperature for
solar-like stars can  also affect the shape of $l_x$.

There is some degeneracy between increasing  $\lambda$ instead of increasing $q$. The former implies a stronger dependence of $l_x$ on $\Theta$  (as discussed at the end of Sec 2.), and thus a stronger dependence of $l_x$ on rotation and time in the unsaturated rotation regime.  Eq. (\ref{lx}) shows that $l_x \propto {\tilde Ro}^{2\over 1-\lambda}$ and observations \citep{Wright2011} then require $0\le \lambda \le 1/3$.
\cite{Blackman2015} showed that $\lambda=1/3$ seems to match the  shape
of the overall averaged $l_x({\tilde Ro})$ curve averaged over stars of different masses and
ages compared to $\lambda=0$.  The higher value of $\lambda$ raises $l_x$ in the saturated regime, and steepens it  in the unsaturated regime.  However for a single star such as the sun, the saturated $l_x$ 
could be lower than the average over a potpourri of late-type stars. More observational constraints on $\Theta$ are needed. The generic effect of changing $\lambda$ is captured in Fig. (\ref{fig2})ab.


 \subsection{Effect of varying $T_{x\odot}$: prediction of  increased spread in $l_x$ with stellar youth: }

\begin{figure}
{\includegraphics[width=0.9\columnwidth]{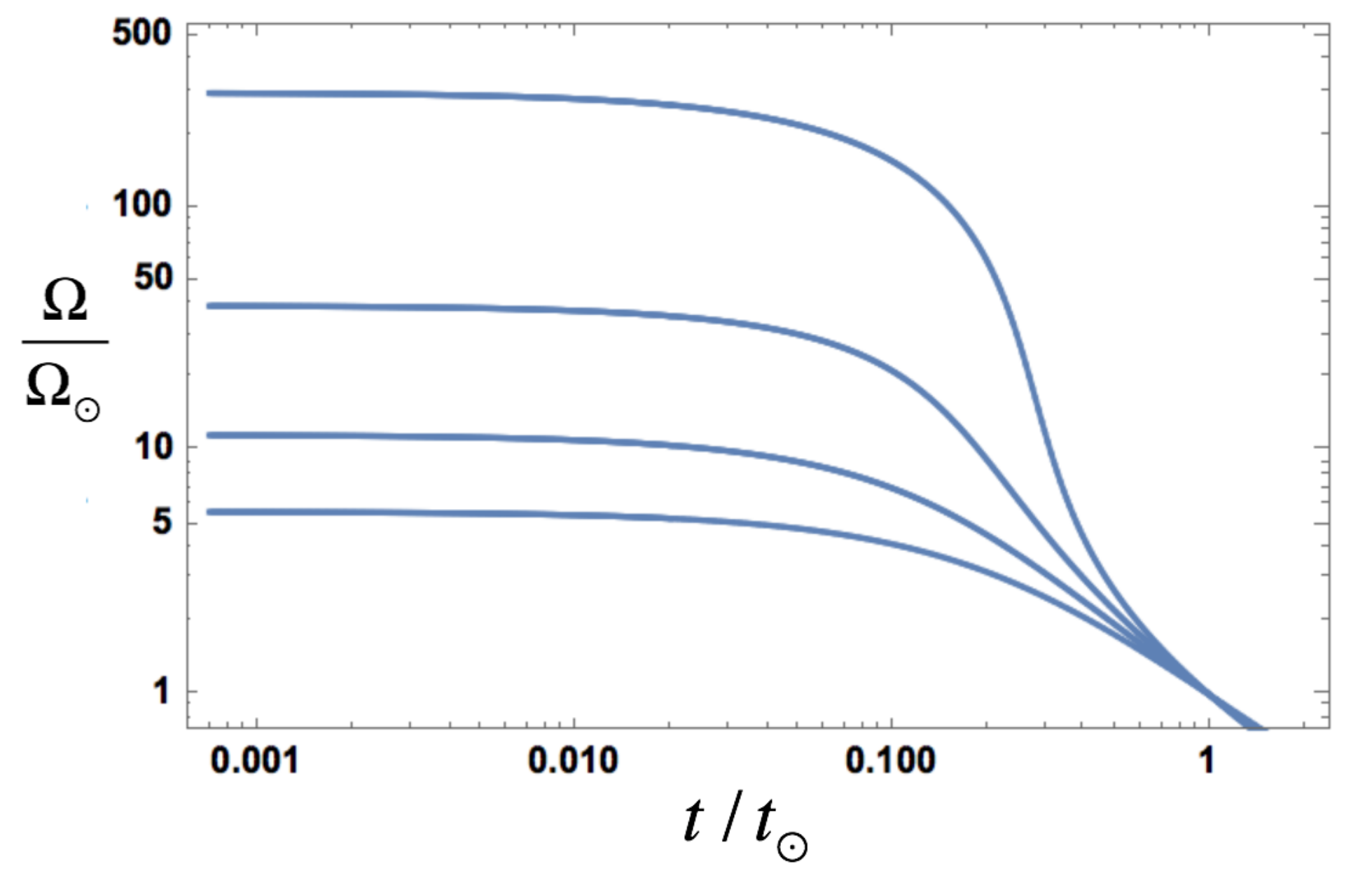}}
{\includegraphics[width=0.9\columnwidth]{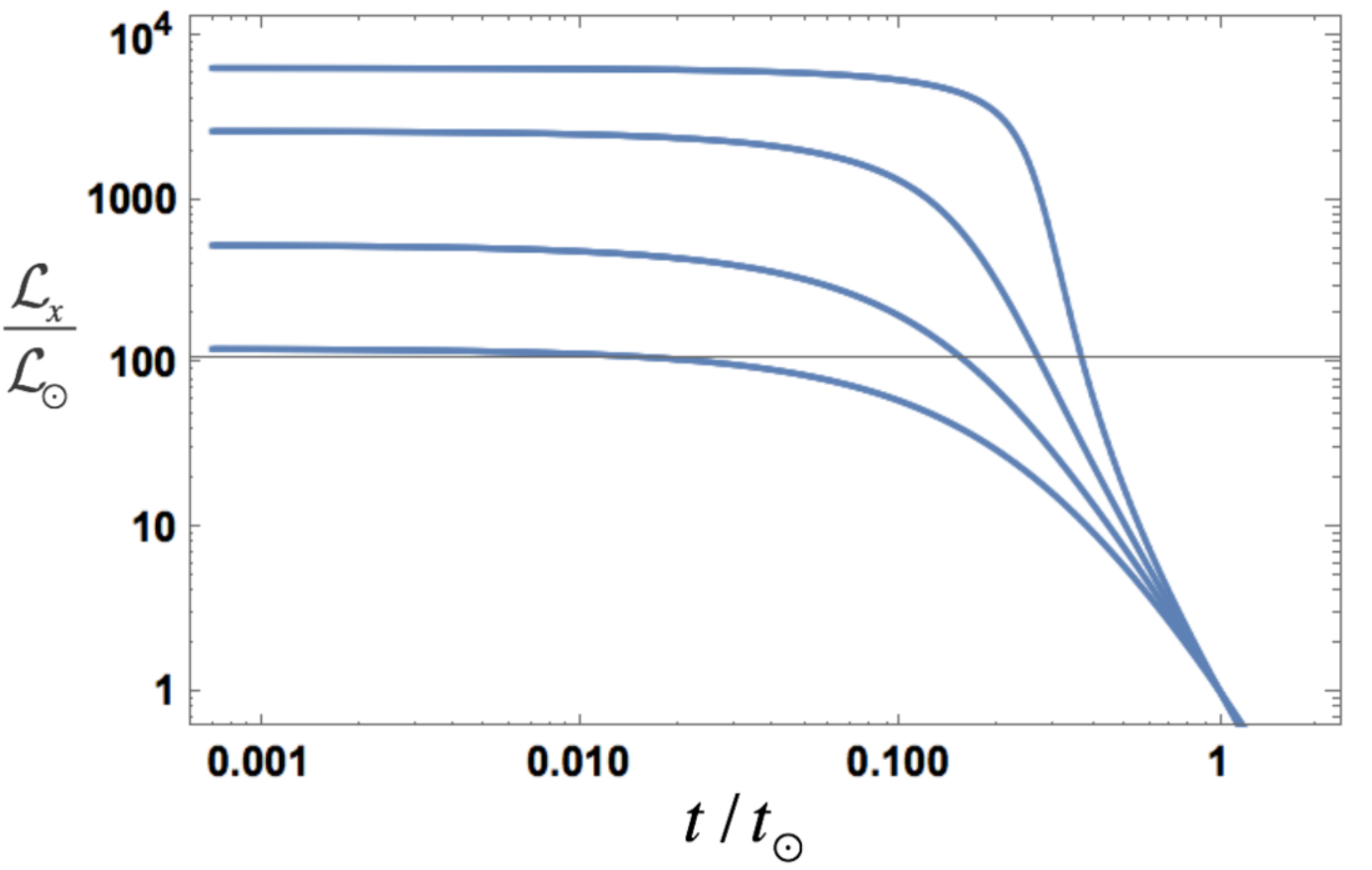}}
 \caption{Our solutions for $\omega_*$ (panel a) 
 and $l_x$  (panel b)
  predict a  broader spread in these quantities at early times 
  compared to late times  if we make small changes in the 
coronal equilibrium temperature normalization for the sun at the current age.
 For all curves,   $\lambda =1/3$, $q=1.2$,  but from bottom to top in each panel 
 we have used  respectively
   ${\tilde T}_{0,\odot} = {1\over 1.6}, {1\over 2.0}, {1\over 2.2}, {1\over 2.4}$.
   The  choice ${\tilde T}_{0,\odot}=1/2$ corresponds to $T_{x,\odot}=1.5 \times 10^6$K and
   is that which was used previously in Section 4.}
 \label{fig5spread}
\end{figure}

One other important implication of our calculations is the 
prediction that the spread in $\omega$ and $l_x$ should increase with decreasing age for a given stellar mass if we allow  small changes to the coronal equilibrium temperature at the solar age.  
This is shown in Fig \ref{fig5spread} and
agrees with the trends reported in \cite{Gallet2013}
The bottom and top curves in each panel of Fig \ref{fig5spread} correspond to the 
extremes ${\tilde T}_{0\odot} = {1\over 1.6}$ and  ${\tilde T}_{0\odot} ={1\over 2.4}$ respectively.
These extremes differ by less than a factor of 2, and are smaller than the reported variation in the Sun's corona temperature \citep[e.g.][]{Johnstone2015}. 

The sensitive dependence of $\omega_*$ and $l_x$ to ${\tilde T}_{0\odot}$ at early times 
can be traced   to the fact that in Eq. (\ref{dotm}), the exponent 
is such that $l_x$ depletes more rapidly toward $l_x\sim 1$ 
for smaller $T_{0\odot}$.  Then when coupled to the time dependence,
 $l_x$ increases with $\omega_*$  in the unsaturated regime, both of these two quantites
 have larger maxima in the saturated regime at early times to arrive at the solar values
 at the sun's age.

Note also that in Eq. \ref{omegadot},  
the ratio of  $b_r^2(t) /u_A(t)$ is less sensitive to $\omega_*(t)$ in the unsaturated
regime than is $\omega_*(t)^{-1/2}$. Thus $\omega_*(t)$
evolves more sharply with the ideal Skumanich law 
but may still produce an "apparent" correspondence
(see Fig \ref{fig2})  if error bars on rotation periods are  with a factor of 2 (highlighting the need for  more data).   
From Eqs. (\ref{2.42}) and (\ref{lx}), we see that
$l_x \propto  \omega_* ^{2\over 1-\lambda}\sim \omega_*^3$ for $\lambda =1/3$
in the unsaturated regime so that the time evolution  of  $l_x$   is more sensitive to $\omega_*$ than $\omega_*$ itself,   explaining why the spread from the top curve to the bottom curve is larger for luminosity than for  $\omega_*$

\section{Conclusion}

To solve for the coupled main sequence stellar evolution of x-ray luminosity, rotation, mass loss rate, and mean magnetic field strength,  we have constructed a simple model combining an isothermal Parker wind 
with sourcing of the needed thermal energy by dynamo generated magnetic fields.
The dynamo produced fields are estimated  from
 a modern  saturation paradigm  based on  magnetic helicity evolution and 
   a physically motivated replacement of the convection time in the dynamo coefficients by  the shear time for fast rotators.  The latter had been previously used in an effort to explain the $l_x(Ro)$ behaviour of late-type stars  but without the  time evolution.  The division of the magnetic sourcing into radiation vs.
   wind is determined by the consequences of assuming energy balance between heating, 
    cooling and mass loss in the corona on time scales short compared to the the spin down.
    We find that ${\mathcal L}_X \sim{\mathcal L}_W$ for the appropriate range of coronal 
    temperatures, masses and radii of early-type stars.
   
The time dependent solutions we obtain exhibit broad-brush  agreement with  observational trends. This includes  time evolution of x-ray luminosities,  magnetic field, and rotation period,
and the approximate  scaling  of mass loss rate with x-ray luminosity.  
We have limited ourselves to a  minimalist holistic time-dependent model as  a framework for more
detailed models which  seems justified,  given  the previous lack of unification of the aforementioned pieces in single theory and the need for more observational data. 

We  have primarily  focused the time evolution of stellar properties of stars younger than the present sun up  to the present day solar  values 
assuming that over the course of the main sequence evolution, the mass and radius  are fixed. The  same approach for stars of other initial masses and  radii normalized to any currently measured fidiucial values would be of interest for future work. Then a population of curves for a range of stars could be generated and averaged properties across stars could be determined.  We have  shown that a population of solar-type stars whose solar-age coronal equilibrium  temperature deviates by less than a factor of two, would produce a broad spread of x-ray luminosities and rotation periods
in youth, converging to a narrow range in old age,  seemingly consistent with observations \citep{Gallet2013}. 

For stars older than the sun, our model suggests that conductive losses  are   relatively more important  than wind losses than for  younger stars (compare regimes I and II of Fig. \ref{fig1}b).
This would  in turn imply a reduction in angular momentum loss, possibly helping to explain the
flattening in the observed period-age relation for old stars \citep{vanSaders2016}.
Exploring  "regime I"  further, and the  implications for angular momentum loss 
are topics for future work.

More detailed extensions of our framework
could include  deviations from spherical symmetry for the wind solution,  
more explicit modeling of closed and open fraction of field lines, more detailed models of mass loss \citep[e.g.][]{Suzuki2013} and spin-down torques \citep[e.g.][]{Matt2015}; 
inclusion of  magnetic contributions to the wind radial momentum equation for fast rotators, a more detailed time dependent dynamo solution, and time dependent modeling of convection and internal stellar structure. The latter may help predict/constrain the free parameter $\lambda$, which may also
be extractable empirically now from Kepler data \citep{McQuillan2014}.
Understanding the mechanisms and dynamics of   angular momentum coupling and decoupling between core and envelope may also  particularly important \citep{Gallet2013,Amard2016}.
Further extensions to include disk-star interactions might be further desirable to extend the framework to pre-main sequence evolution that would predict
the "initial" conditions for the post-disk main sequence evolution.

With each additional  complexity   also comes additional caution that the ingredients that add to the complexity may themselves not be well understood. For example,  how dynamos  work in stars,  where in the star the field is actually amplified and anchored, and how the interior dynamo and anisotropic convection evolve with time and connect to the coronal field remain  matters of active research, even for steady state conditions. 
In this sense, there is value both in minimalist approaches as well as  detailed modeling.


\section*{Acknowledgments} 
We thank  the referee for a helpful and insightful report, and E. Mamajek for related discussions. 
 EB acknowledges  grant support from 
 NSF-AST-1109285, HST-AR-13916.002, a Simons  Fellowship, and the IBM-Einstein Fellowship Fund  while at IAS in 2014-2015. JEO acknowledges support by NASA through Hubble Fellowship grant HST-HF2-51346.001-A awarded by the Space Telescope Science Institute, which is operated by the Association of Universities for Research in Astronomy, Inc., for NASA, under contract NAS 5-26555.

\bibliographystyle{mn2e}
\bibliography{spindownbib}

\end{document}